\documentclass[twocolumn]{aastex7}

\usepackage{graphicx}

\graphicspath{{.}}  

\usepackage{amsmath}
\usepackage{amssymb}

\usepackage[usenames,dvipsnames]{xcolor}

\usepackage{amstext}
\usepackage{mathtools}
\usepackage{siunitx}
\usepackage{physics}
\usepackage{xspace}
\usepackage{multirow}
\usepackage{makecell}
\usepackage{booktabs}

\usepackage{tabularx}

\usepackage{newtxtext,newtxmath}

\usepackage{cleveref}

\makeatletter
\pretocmd\@sect{\def\@currentcounter{#1}}{}{\fail}
\makeatother

\DeclareSIUnit{\parsec}{pc}
\DeclareSIUnit{\pc}{pc}
\DeclareSIUnit{\year}{yr}
\DeclareSIUnit{\Gyr}{Gyr}
\DeclareSIUnit{\Myr}{Myr}
\DeclareSIUnit{\dex}{dex}
\newcommand*{\Msun}{\ensuremath{\mathrm{M}_{\odot}}\xspace}

\newcommand*{\limepy}{\textsc{limepy}\xspace}
\newcommand*{\emacss}{\textsc{emacss}\xspace}
\newcommand*{\GCfit}{\textsc{GCfit}\xspace}
\newcommand*{\ssptools}{\textsc{SSPtools}\xspace}
\newcommand*{\clusterBH}{\textsc{clusterBH}\xspace}
\newcommand*{\Nbody}{\(N\)-body\xspace}
\newcommand*{\coupled}{\texttt{cBH+limepy}\xspace}

\newcommand*\NGC[1]{NGC\thinspace{#1}}

\newcommand*{\omegacen}{\(\omega\)\thinspace{Cen}\xspace}

\newcommand*\chem[1]{\ensuremath{\mathrm{#1}}}
\newcommand*{\FeH}{\ensuremath{[\chem{Fe}/\chem{H}]}}

\newcommand*{\rhi}{\ensuremath{r_{\mathrm{h},0}}\xspace}
\newcommand*{\rhoh}{\ensuremath{\rho_{\mathrm{h}}}\xspace}
\newcommand*{\rhohi}{\ensuremath{\rho_{\mathrm{h},0}}\xspace}
\newcommand*{\Mi}{\ensuremath{M_{0}}\xspace}

\newcommand*{\fbhi}{\ensuremath{f_{\mathrm{BH},0}}\xspace}
\newcommand*{\fbh}{\ensuremath{f_{\mathrm{BH}}}\xspace}

\newcommand*{\Mbh}{\ensuremath{M_{\mathrm{BH}}}\xspace}

\newcommand*{\mbh}{\ensuremath{m_{\mathrm{BH}}}\xspace}

\newcommand*{\fret}{\ensuremath{f_{\mathrm{ret}}}\xspace}

\newcommand*{\vesci}{\ensuremath{v_{\mathrm{esc},0}}\xspace}

\newcommand*{\fkick}{\ensuremath{f_{\mathrm{k}}}\xspace}
\newcommand*{\RGeff}{\ensuremath{R^\prime_{\mathrm{G}}}\xspace}

\newcommand*{\rahat}{\ensuremath{\hat{r}_{\mathrm{a}}}}

\newcommand*{\rh}{\ensuremath{r_{\mathrm{h}}}}

\newcommand*{\rc}{\ensuremath{r_{\mathrm{c}}}}

\newcommand*{\Gaia}{\textit{Gaia}\xspace}
\newcommand*{\HST}{\textit{HST}\xspace}

\defcitealias{Fronimos2026}{F26}
\newcommand*{\cbhpaper}{\citetalias{Fronimos2026}\xspace}

\defcitealias{Dickson2023}{D23}
\newcommand*{\MMpaperI}{\citetalias{Dickson2023}\xspace}

\defcitealias{Dickson2024}{D24}
\newcommand*{\MMpaperII}{\citetalias{Dickson2024}\xspace}

\defcitealias{Dickson2023,Dickson2024}{Dboth}
\newcommand*{\MMpapers}{\citetalias{Dickson2023,Dickson2024}\xspace}

\begin{document}


    \title{
        Fast Dynamical Modelling of Milky Way Globular Clusters --
        I. Implications for Initial Cluster Densities
    }
    \shorttitle{
        Fast Dynamical Modelling of MW GCs
    }


    \author[orcid=0000-0002-6865-2369,gname=Nolan,sname=Dickson]{Nolan Dickson}
    \affiliation{Department of Astronomy and Physics, Saint Mary’s University, 923 Robie Street, Halifax, NS B3H 3C3, Canada}
    \email[show]{nolan.dickson@smu.ca}

    \author[orcid=0000-0003-2927-5465,gname=H\'enault-Brunet,sname=Vincent]{Vincent H\'enault-Brunet}
    \affiliation{Department of Astronomy and Physics, Saint Mary’s University, 923 Robie Street, Halifax, NS B3H 3C3, Canada}
    \email{vincent.henault@smu.ca}

    \author[orcid=0000-0003-4158-5044,gname='Fronimos Pouliasis',sname=Fotios]{Fotios Fronimos Pouliasis}
    \affiliation{Institut de Ciències del Cosmos (ICCUB), Universitat de Barcelona (UB), c. Martí i Franqués, 1, 08028 Barcelona, Spain}
    \email{fotisfronimos@icc.ub.edu}

    \author[orcid=0000-0002-9716-1868,gname=Gieles,sname=Mark]{Mark Gieles}
    \affiliation{Institut de Ciències del Cosmos (ICCUB), Universitat de Barcelona (UB), c. Martí i Franqués, 1, 08028 Barcelona, Spain}
    \affiliation{ICREA, Pg. Llu\'{i}s Companys 23, E08010 Barcelona, Spain}
    \affiliation{Institut d'Estudis Espacials de Catalunya (IEEC), Edifici RDIT, Campus UPC, 08860 Castelldefels (Barcelona), Spain}
    \email{mgieles@icc.ub.edu}

    \author[orcid=0000-0002-7489-5244,gname=Smith,sname=Peter]{Peter J. Smith}
    \affiliation{Max Planck Institute for Astronomy, K\"onigstuhl 17, D-69117 Heidelberg, Germany}
    \affiliation{Department of Physics and Astronomy, University of Heidelberg, Im Neuenheimer Feld 226, D-69120 Heidelberg, Germany}
    \email{pesmith@mpia.de}


    \date{Accepted XXX. Received YYY; in original form ZZZ}



    \begin{abstract}

    We infer the initial conditions of Milky Way (MW) globular clusters
    (GCs) from present-day observations, through
    the coupling of recently updated rapid cluster evolution models
    with multimass equilibrium models.
    This novel method is validated by fitting to simulated observations
    of a large grid of star-by-star Monte Carlo models, demonstrating that we
    are able to recover cluster properties like the total mass, half-mass
    radius/density and black hole (BH) mass fraction, both
    initially and at the present day, across a large region of parameter space.
    We apply this framework to a sample of 40 MW GCs,
    fitting to a suite of observed radial profiles of number densities,
    proper motions, line-of-sight velocities and stellar mass functions.
    From these fits we infer a distribution of initial half-mass densities
    with a median and \(1\sigma\) width, across our sample, of
    \(\rhohi = 10^{6.4\pm0.9}\,\unit{\Msun \ \pc^{-3}}\),
    higher than what is found for young massive clusters in the Local
    Universe and in line with young clusters at high redshift.
    We also find stellar initial mass functions that are bottom-light in
    comparison to canonical prescriptions,
    and relatively small present-day BH mass fractions (\(\lesssim 1.5\%\)).
    We discuss the implications of these initial cluster densities for
    observations of high-redshift proto-GCs, binary BH merger rates and
    intermediate-mass BHs (IMBHs) in GCs.
    Finally, we quantify how these densities may depend on
    assumptions typically made surrounding BH formation and natal kicks.

    \end{abstract}




\section{Introduction}\label{sec:introduction}



    Massive globular star clusters (GCs) provide an important fossil record
    for the study of large and small structures in the Universe, and are
    key for understanding a number of important cosmic processes
    such as star formation, galaxy assembly, black hole (BH) growth and
    gravitational waves (GWs).


    Despite their importance, and ubiquity in galaxies like our own Milky Way
    (MW), the precise origins and initial conditions of the ancient GCs we
    see today remains a much studied, but inconclusive, question
    \citep[see e.g.][]{Forbes2018,Kruijssen2026}.
    The formation and early evolution of clusters are essential to
    the progression of their entire lifetimes, and influence the possible
    growth of intermediate-mass BHs (IMBHs) \citep[e.g.][]{PortegiesZwart2002,
    Fujii2024} and multiple populations \citep{Bastian2018}.
    However, we cannot easily observe these conditions directly, as nearby GCs
    are universally ancient systems, and differences with
    observed young massive clusters (YMCs) sow uncertainty about whether the
    young clusters in the local Universe will evolve one day into GCs as we
    see at the present day, or if clusters formed differently in the ancient
    Universe than today \citep{PortegiesZwart2010,Krumholz2019}.

    Modern high-resolution simulations of GC formation attempt to reproduce
    observed cluster populations through the inclusion of a number of physical
    processes, including star formation, stellar and binary evolution,
    and collisional dynamics, across a variety of interstellar environments
    \citep[e.g.][]{Cournoyer-Cloutier2024,Polak2024,Reina-Campos2025,
    Lahen2025a,Lahen2025b,Williams2025}.
    However, the self-consistent modelling of cluster formation and evolution
    remains too computationally expensive to enable the exploration of large
    regions of possible parameter space.

    In recent years, with the advent of JWST, observations of strongly
    gravitationally lensed galaxies have enabled the discovery of young and
    massive candidate proto-GCs at high redshifts
    \citep[e.g.][]{Vanzella2023,Claeyssens2023,Adamo2024,Claeyssens2026}.
    Though there remain significant uncertainties in the derivation of the
    physical properties of these systems, and it is still not entirely clear
    if these proto-GCs are destined to evolve into the MW-like GC populations
    we see today.


    On the opposite end of cluster lifetimes, at the present-day, we have
    access to extensive observations of GCs within the MW from
    ground and space-based facilities \citep[e.g.][]{Watkins2015,Kamann2018,
    Libralato2022,GaiaCollaboration2023,Libralato2024}.
    This data has enabled the detailed modelling of a large number of GCs
    through various approaches \citep[e.g.][]{Zocchi2019,Henault-Brunet2019,
    Henault-Brunet2020,Rui2021b,Vitral2022,Vitral2023}.
    These efforts have been used to explore the structure, kinematics,
    chemistry, and stellar and remnant populations of these systems.
    Of particular note, many recent studies have explored the populations
    of BHs which may exist today within most GCs, through both direct
    observations \citep[e.g.][]{Strader2012,Paduano2024,Giesers2018,
    Giesers2019,GaiaCollaboration2024,Whitaker2026}, and indirect modelling
    \citep[e.g.][]{Weatherford2018,Weatherford2020,Askar2018}.

    In \citet[][hereafter \MMpapers]{Dickson2023,Dickson2024}, we harnessed a
    variety of observational datasets in order to fit multimass
    equilibrium models to a large sample of the most well studied MW GCs.
    These models were used to infer the composition and distribution of
    luminous and dark mass within the clusters, and in particular to place
    constraints on their stellar initial mass functions (IMF) and BH
    populations, finding (alongside \citealt{Baumgardt2023}) a non-canonical,
    bottom-light IMF and relatively small (but typically non-zero) total BH
    masses.


    These observations and models of MW clusters at the present day,
    \(\sim\SI{12}{\Gyr}\) after their birth, represent a key boundary condition
    in our attempts to constrain the formation and long-term evolution of GCs.
    However, bridging this gap between the initial and current conditions
    of clusters is not straightforward, as the clusters undergo Gyrs of
    evolution driven by internal and external mechanisms impacting their mass
    loss and size evolution.
    Additionally, the evolution may be strongly impacted by a number of
    smaller-scale physical processes which remain uncertain, such as
    the stellar IMF, or the supernovae natal kicks received by BHs upon their
    formation \citep[e.g.][]{Popov2025}.



    The long-term dynamical evolution of GCs is driven by a
    tendency towards a state of unstable thermal equilibrium,
    and the corresponding flow of heat from the core
    to satisfy the energy demands of the system \citep{Henon1961}.
    This evolution is the emergent
    result of a number of physical processes, such as two-body relaxation,
    stellar evolution and evaporation in the tidal field of the host galaxy
    \citep{Spitzer1987,Heggie2003}.
    A great deal of research has approached this complex evolution through
    direct \Nbody integration methods. However, this comes with significant
    computational costs, and only recently has it become possible to simulate
    clusters of \(>10^6\) stars
    \citep[e.g.][]{Wang2016,ArcaSedda2024,Bianchini2026}.
    \citet{Henon1971a,Henon1971b} style Monte Carlo dynamical models, such as
    Cluster Monte Carlo \citep[CMC;][]{Rodriguez2022} and Monte Carlo Cluster
    Simulator \citep[MOCCA;][]{Hypki2013,Giersz2013}, offer significantly lower
    computational costs and are used to simulate clusters of up to
    \(10^7\) stars \citep[e.g.][]{Mai2026}.
    However, even with these faster methods, the computational costs are still
    so large that it is only possible to compute limited grids of large-\(N\)
    models, and a complete exploration of realistic parameter space remains
    infeasible.


    If we are willing to restrict our analysis to the bulk properties and
    overall stellar and remnant populations within a cluster, rather than
    examining the individual stars and remnants in detail, the fastest approach
    to modelling the dynamical evolution of GCs can be offered by
    semi-analytical models, with simple, physically motivated prescriptions
    describing the impacts of various relevant physical processes on
    cluster quantities like mass and radius over time.

    \citet{Gieles2011} presented an expression for the evolution of the
    mass and size of a cluster within a tidal field based on the unification
    of the original models of Henon for isolated \citep{Henon1965} and tidally
    limited clusters \citep{Henon1961}.
    \citet{Henon1975} demonstrated that the heat flowing from the core of
    a cluster is controlled by the two-body relaxation demands of the rest of
    the system,
    while \citet{Breen2013} showed that this energy could be produced
    by a population of binary BHs (BBH), segregated to the core, through
    interactions causing the formation and hardening of the binaries, a process
    known as BH-burning.
    Based on these recipes, \citet{Antonini2020a,Antonini2020b,Antonini2023}
    developed the \clusterBH cluster evolution models,
    which self-consistently follow the evolution of both the cluster and
    its internal BH subsystem.
    Analogous models have also been constructed recently following similar
    prescriptions, such as
    \emacss \citep{Alexander2012,Gieles2014,Alexander2014}, \textsc{FastCluster}
    \citep{Mapelli2021} and \textsc{Rapster} \citep{Kritos2024}.


    Recently, the \clusterBH models were expanded upon in
    \citet[][hereafter \cbhpaper]{Fronimos2026} to account for many other
    relevant physical processes, such as the effects of external tidal fields,
    a range of cluster metallicities, and different IMFs.
    These semi-analytical models were calibrated against a large grid of
    CMC cluster models \citep{Kremer2020}, covering a range of initial
    conditions, and were able to reproduce the evolution of these star-by-star
    models to within about 10 per cent, and their BBH merger rate to within
    about 20 per cent.
    The speed, flexibility and accuracy of these updated models provides a
    useful tool enabling the exploration of a number of open questions
    surrounding topics such as BH growth
    (\citealt{Chattopadhyay2026}, Marín Pina et al., in prep.), GW
    population synthesis (F. Fronimos Pouliasis et al., in prep.),
    GC tidal streams (F. Fronimos Pouliasis et al., in prep.) or,
    as we examine here, the initial conditions of GCs.



    In this work, we combine these fast evolutionary models with the
    multimass equilibrium models previously used in \MMpapers,
    allowing us to bridge the gap between the present-day observations
    and the unknown early conditions of a large sample of MW GCs.
    We infer in detail the distributions of stars and remnants
    in these clusters today by fitting on various observables, and working
    backwards through the clusters' evolution, place constraints on
    their initial conditions.


    In \Cref{sec:methods}, we describe the evolutionary \clusterBH models,
    the multimass \limepy equilibrium models, and how they are coupled together,
    as well as the observational datasets and model-fitting procedures used.
    In \Cref{sec:validation}, we test the ability of these models to reproduce
    the evolution of realistic clusters by fitting our models to simulated
    observations extracted from star-by-star dynamical models.
    Results for the fits to our MW cluster sample are then presented and
    discussed in \Cref{sec:real_fitting}, where we examine in more detail the
    inferred distributions of BH mass retained to the present day
    (\Cref{sub:black_hole_populations}), the stellar IMF
    (\Cref{sub:initial_mass_function}) and the cluster initial
    conditions (\Cref{sub:initial_conditions}).
    Finally, in \Cref{sec:discussion}, we discuss the implications of our
    results on cluster formation, IMBH growth, and GW rates,
    before concluding in \Cref{sec:conclusions}.



\section{Methods}\label{sec:methods}

\subsection{\clusterBH Evolutionary Models}
\label{sub:evolutionary_models}


    To model the bulk evolution of GCs and their BH subsystems
    from initial conditions to the present day, we use the latest version of
    the \clusterBH fast evolutionary models, as described in \cbhpaper
    \footnote{Available at \url{https://github.com/cBHBd/cBHBd}. Note that
    while \clusterBH is packaged alongside the BHB population synthesis models
    \textsc{BHBdynamics}, here \clusterBH is used in isolation.}.

    In short, \clusterBH functions by considering the changes in the overall
    energy, total mass, mass in BHs and (through the assumption of virial
    equilibrium) half-mass radius of a cluster, as a result of a few major
    physical mechanisms, including two-body relaxation, stellar evolution and
    tidal evaporation to a host galaxy.


    Two-body relaxation contributes to the cluster energy evolution after the
    formation of the first BBH in the cluster core, which we assume begins at
    the ``core collapse time'' \(t_{cc}\), after which the energy production
    in the core and the energy requirements of two-body relaxation in the rest
    of the cluster balance one another.
    The relaxation timescale, and thus the rate of change of the total cluster
    energy (excluding the negative energy locked in multiples), is
    itself related to the relative masses and radii of the stellar and BH
    populations within the cluster, and is thus dependent on the cluster
    metallicity and stellar IMF.
    The BH-burning process which generates the energy in the core of the cluster
    also results in ejections of BHs from the cluster. \clusterBH
    follows the prescriptions of \citet{Breen2013} while in the balanced phase
    of evolution, but with an extra reduction to the BH ejection efficiency at
    later stages, to match the behaviour observed in \Nbody models.


    Stellar evolution mass loss, through stellar winds and supernovae,
    contributes continuously to the expansion of the cluster over its lifetime,
    after some timescale \(t_{\mathrm{sev}}\) marking the start of stellar
    evolution. It is most dominant at early times, as the most massive
    stars evolve and die.
    While \clusterBH is essentially a two-component model, consisting of BHs
    and all other non-BH objects, we note that the formation and
    retention of all non-BH stellar remnants are accounted for in the
    ``stars'' component and stellar-evolution driven mass loss mechanisms.


    Finally, the presence of an external tidal field from a host galaxy like the
    MW causes the gradual evaporation of stars
    (preferentially low-mass) over the tidal boundary of the cluster.
    This tidal boundary is itself dependent on the strength of the external
    galactic potential, and thus the location of the cluster within the galaxy.
    In this work, we assume a static potential for the MW in the form
    of a singular isothermal sphere with a circular velocity of
    \SI{220}{\kilo\metre\per\second}. An effective galactocentric radius for
    each cluster must be determined based on its orbit. This is discussed in
    more detail in \Cref{ssub:cluster_data}.

    \cbhpaper introduced 7 model parameters and a number of constants which
    define the various processes in \clusterBH, and calibrated the models
    through comparison with a large sample of CMC dynamical models
    \citep{Kremer2020,Rodriguez2022}.
    In this work, we use the default values of each of these parameters, which
    correspond to the median best-fitting values found in \cbhpaper (see their
    Table 2).

\subsection{Multimass \limepy Equilibrium Models}
\label{sub:equilibrium_models}


    To model the present-day phase-space distribution of the GCs we use
    the \limepy multimass distribution-function (DF) based models
    \citep{Gieles2015}, through the \GCfit Python package, as described in
    \MMpapers
    \footnote{Available at \url{https://github.com/mgieles/limepy} and
    \url{https://github.com/nmdickson/GCfit}.}.
    The multimass version of these DF based models uses a set of discrete
    individual mass bins (described by the total (\(M_j\)) and mean (\(m_j\))
    masses of each bin), representing a spectrum of stellar and remnant masses.
    The total DF is then defined as the sum of component DFs for each mass bin.
    These individual mass bins are required in order to account for important
    dynamical effects such as mass segregation, and to track specific types of
    objects, such as BHs.

    The multimass models used here are defined by a number of
    free parameters, which dictate the physical solution of the \limepy DF.
    The central concentration of the model is defined by the (dimensionless)
    central potential parameter \(\hat{\phi}_0\).
    The sharpness of the energy truncation near the tidal radius of the
    cluster, mimicking the effects of the host galaxy's tides, is given by
    the parameter \(g\), with lower values indicating a more abrupt truncation.
    For certain values of \(g\), \limepy recovers some well-known families of
    models, such as the multimass King model
    \citep[for \(g=1\);][]{Michie1962,King1966}.
    The size and mass scales of the model are represented by the (present-day)
    total cluster mass \(M\) and 3D half-mass radius \(\rh\) parameters.
    The level of velocity anisotropy present in the model is dictated by the
    (dimensionless) anisotropy radius parameter \(\rahat\) (in units of the
    King radius \(r_0\)), defining the
    radius at which the isotropic model core becomes radially anisotropic,
    before returning again to isotropy at the truncation radius.

    The exact definition of \(\hat{\phi}_0\) and \(\rahat\) depends on the
    definition of the mean mass within \limepy, and, in contrast to \MMpapers,
    here we choose to adopt the central mean mass, as defined in
    \citet{Gieles2015}, which we find to enable more stable computation of
    models containing many BHs
    \footnote{It is straightforward to convert the values of
    \(\hat{\phi}_0\) and \(\rahat\) found under either the global or central
    mean mass definitions from one to the other, for comparison with other
    literature results, using equations 8 and 9 of \citet{Peuten2017}.}.

    \subsubsection{Updates to Energy Equipartition Prescriptions}

    Within the multimass version of the \limepy DF, the trend towards
    energy equipartition between objects of different masses, and the effects
    of mass segregation
    \citep{Gieles2015, Peuten2017, Henault-Brunet2019}, are
    captured by a mass-dependent velocity scale \(s_j\), which is
    parametrized by the free parameter \(\delta\), such
    that \(s_j^2 \propto m_j^{-2\delta}\). Higher values of \(\delta\) represent
    increasingly mass segregated models, with \(\delta=1/2\) representing a
    cluster that would be in equipartition in the limit
    \(\hat{\phi}_0 \to \infty\).

    However, this implies a shared equipartition relationship across all
    masses, including both stars and BHs.
    In reality, we expect BHs, which segregate very rapidly to form their own
    subsystem in the cluster core,
    to partly ``decouple'' from the rest of the stars.
    These central BHs should suppress the mass segregation of visible stars
    by dynamically heating the more massive stars and preventing their
    stratification to the core \citep{Alessandrini2016,Peuten2017}.
    However, a single shared equipartition parameter (\(\delta\)) is unable to
    simultaneously capture both the segregation of BHs and the suppression of
    the equipartition among the rest of the stars at the same time
    Therefore, in this work, we introduce a modified prescription
    for energy equipartition in \limepy.

    To allow the BHs to partly decouple from the rest of the cluster, we instead
    compute the mass-dependent velocity scale parameter as:
    \begin{equation}
        s_j^2 \propto
        \begin{cases}
            m_j^{-2\delta}, & m_j < m_{\mathrm{lim}} \\
            \zeta\ m_j^{-2\delta}, & m_j \geq m_{\mathrm{lim}}
        \end{cases}
    \end{equation}
    where \(\zeta\) is a new free parameter, between 0 and 1, which allows the
    velocity scale to be reduced for all masses above a certain limiting value,
    which we take here to be \(m_{\mathrm{lim}}=\SI{3}{\Msun}\), to capture
    only the BHs.
    In tandem with this new prescription, we also now allow the
    anisotropy radius to be mass-dependent, by allowing the parameter
    \(\eta\) (such that \(r_{{\mathrm{a}},j}\propto r_{\mathrm{a}}\ m_j^\eta\))
    to vary freely, whereas in \MMpaperI this was fixed to \(\eta=0\).

    These changes allow the BHs in our models to be much more centrally
    concentrated, consistent with star-by-star GC models
    \citep[e.g.][]{Breen2013}, and impacting the effects of the BH
    population on certain observable quantities, like the velocity dispersion.

\subsection{Coupling \clusterBH and \limepy}
\label{sub:coupled_models}


    The two models, \clusterBH (\Cref{sub:evolutionary_models}) and
    \limepy (\Cref{sub:equilibrium_models}), are coupled
    together in this work (hereafter referred to as \coupled), allowing us to
    infer the initial conditions of real GCs from their present-day properties
    alone.
    To reflect this, in \coupled the free parameters of total mass
    (\(M\)) and half-mass radius (\rh) are replaced with their initial (time 0)
    counterparts (\Mi, \rhi).


    The IMF, necessary to determine both the
    initial BH mass function (e.g. total and average BH mass) within \clusterBH
    and the present-day stellar and remnant mass bins within the
    multimass \limepy models, is implemented as a broken three-component power
    law, in keeping with the parametrization of canonical IMFs
    \citep[e.g.][]{Kroupa2001}. The three power-law slopes are given by the
    parameters \(\alpha_1,\,\alpha_2,\,\alpha_3\) (with break masses at 0.5 and
    1 \Msun and bounded between 0.08 and 150 \Msun).
    As there is increasing evidence that the low-mass IMF of MW GCs
    does not follow the commonly assumed \citet{Kroupa2001} IMF
    \citep{Baumgardt2023,Dickson2023},
    we wish to explore various IMFs with our models, and thus we let
    the \(\alpha_1\) and \(\alpha_2\) slopes vary freely.


    These initial quantities are used to evolve, through clusterBH, the bulk
    quantities of the cluster over its lifetime, to its present-day age. The
    present-day total mass, half-mass radius and BH mass fraction are then used,
    alongside the other structural free parameters described in
    \Cref{sub:equilibrium_models}, to compute the multimass equilibrium \limepy
    DF based model.
    To bridge the two models, we first determine from these
    quantities the present-day mass function (PDMF), which describes the mass
    bins in our multimass models, through a separate prescription.
    As described in \MMpaperI, we do so using the mass
    function evolution algorithm implemented in the \ssptools library
    \footnote{Available at https://github.com/SMU-clusters/ssptools.}.

    In short, this algorithm determines the number of stars which evolve off the
    main sequence over the lifetime of a cluster based on interpolated
    Dartmouth Stellar Evolution Program models \citep{Dotter2007,Dotter2008},
    and determines the types and masses of the resulting remnants based on
    their initial mass, metallicity and an initial-final mass relation (IFMR).


    The white dwarf (WD) IFMR is interpolated from the MIST 2018 isochrones
    \citep{Dotter2016,Choi2016}.
    All remnants with progenitor masses between the maximum WD progenitor
    mass and the minimum BH progenitor mass are assumed to be neutron stars
    (NS) with a mass of 1.4 \Msun.
    The BH IFMR is interpolated from
    a grid of stellar evolution library (SSE) models of different metallicities,
    using the updated version of SSE presented by \citet{Banerjee2020} and the
    rapid supernova scheme \citep{Fryer2012}.
    The effects of pair-instability supernovae and pulsational
    pair-instability supernovae are implemented according to the
    prescriptions of \citet{Belczynski2016}.


    As these remnants are formed, we mimic the effects of natal kicks causing
    the escape of these objects from the cluster by scaling the final number and
    mass of the created remnants by an ``initial retention fraction''
    \(f_{\mathrm{ret}}\).
    For WDs, we assume \(f_{\mathrm{ret}}=100\) per cent.
    We assume a NS retention fraction of 10 per cent, as is common
    \citep[e.g.][]{Pfahl2002}, though it has also been shown that the results
    of mass modelling are insensitive to the exact NS retention fraction
    due to the low total mass in NS \citep{Henault-Brunet2020}.

    The natal kicks of BHs are modelled in a more in-depth fashion to derive
    a value for \(f_{\mathrm{ret}}\) as a function of the progenitor mass.
    By default, we follow a commonly assumed prescription, beginning by
    assuming the kick velocity is drawn from a Maxwellian distribution,
    with a dispersion of \(\sigma=(1-f_b)\,\SI{265}{\km\per\s}\),
    where the base \SI{265}{\km\per\s} is the dispersion which has been found
    for NS \citep{Hobbs2005}\footnote{Although see also \citet{Disberg2025}.},
    and \(f_b\) is the fallback fraction, the
    fraction of the stellar envelope which falls back onto the BH, which we
    interpolate from the same grid of SSE models as the BH IFMR.
    This fallback fraction is a function of both metallicity and stellar mass,
    and dictates how the kick velocity distribution varies, typically from
    full kicks at low BH masses to zero kicks at the highest masses,
    where BHs form without a supernova explosion
    \citep[direct collapse;][]{Fryer2012}.
    We then compute the fraction of BHs retained in the cluster after the
    kicks by integrating this distribution up to the initial cluster central
    escape velocity, i.e. by evaluating the cumulative distribution function
    (CDF) of the Maxwellian at \vesci:
    \begin{equation}
        \fret(\mbh) = \operatorname{erf}
            \left( \frac{\vesci}{\sqrt{2}\sigma} \right)
            - \sqrt{ \frac{2}{\pi} }\,\frac{\vesci}{\sigma}\,
              \exp\left(\frac{-\vesci^{2}}{2\sigma^2}\right)
    \end{equation}
    where, as given above, \(\sigma\) is a function of the fallback fraction
    and the escape velocity is computed from the initial mass \Mi and
    average half-mass density \rhohi (as in \cbhpaper):
    \begin{equation}
        \vesci \simeq \SI{50}{\km\per\s}\,
            \left(\frac{\Mi}{{10^5}\,\unit{\Msun}}\right)^{1/3}
            \left(\frac{\rhohi}{{10^5}\,\unit{\Msun\pc^{-3}}}\right)^{1/6} .
    \end{equation}

    While this ``canonical'' prescription is commonly used, it is important to
    note that the true natal kick velocity distribution of BHs remains
    uncertain.
    Extensive modelling of BH formation and supernova explosions remains
    difficult and computationally expensive, while the limited examples of both
    direct and indirect observations of the effects of natal kicks on BHs
    provide an incomplete, and at times contradictory, measure of the
    distributions of velocities imparted by kicks
    \citep[e.g.][]{Mandel2016,Repetto2017,Atri2019,Popov2025,Willcox2025a,
    Willcox2025b}.
    Even within the commonly used setup of fallback-modulated Maxwellian
    BH kick distributions, the exact details of the chosen supernova engine can
    result in different expected fallback fraction curves, for example between
    the rapid and delayed convection mechanisms of \citet{Fryer2012}.

    Our flexible population synthesis models allow for
    variations in the magnitude and shape of kick distributions,
    which may have important implications on the required initial conditions
    of clusters. The potential impacts of different BH natal kick assumptions
    and IFMRs are discussed briefly in \Cref{sub:implications_for_bh_physics}
    and will be explored in more detail in a forthcoming work, while in this
    paper we adopt the above canonical prescriptions.


    The binned BH mass function resulting from the creation of all BHs that will
    form from a given IMF and be initially retained in the cluster after natal
    kicks is also used by \clusterBH to determine the initial BH mass fraction
    and track the mean BH mass over time.


    In contrast to \MMpapers, the extra temporal
    information available from \clusterBH also allows us to
    account for and normalize the escape rates of stars and BHs
    within \ssptools.
    To simulate the dynamical ejections of BHs over time from the cores of GCs,
    BHs are removed beginning from the heaviest mass bins (with larger
    gravitational interaction cross-sections) through to the lightest
    \citep[e.g.][]{Morscher2015,Antonini2020a}, until the total BH mass fraction
    over time \(\fbh(t)\) matches that of the \clusterBH model.
    While in \MMpapers we were only able to allow the final BH
    retention fraction to vary freely, the coupling with
    the evolutionary models now provides more realistic, self-consistent
    constraints on the possible dynamical retention history of the BH
    populations.


    The loss of stars, including all non-BH remnants, through dynamical
    ejections and tidal evaporation is modelled by assuming that
    the rate of change of the mass function over time goes as
    \begin{equation}
        \dot{f}(t,m) = -C(t) f(m)h(m),
    \end{equation}
    where \(f(m)\) is the mass function (number of stars) at a given mass \(m\),
    \(C(t)\) is the normalization factor and \(h(m)\) is a function which
    accounts for the preferential loss of low-mass stars within the balanced
    evolution phase \citep{Balbinot2018}, such that
    \begin{equation}
        h(m) = \begin{cases}
            1 - \left(\frac{m}{m_{\mathrm{d}}}\right)^{1/2} &\ ,\ m < m_{\mathrm{d}} \\
            0  & \ ,\ m > m_{\mathrm{d}}
        \end{cases}
    \end{equation}
    where \(m_{\mathrm{d}}\) is the depletion mass, which we fix at 1.2 \Msun,
    based on the typical results of \citet{Lamers2013}.
    Typically, determining the normalization factor, which is set by
    the total stellar mass loss rate, is non-trivial; however this information
    is readily available from \clusterBH, allowing us to compute
    \(C(t)\) directly:
    \begin{equation}
        C(t) = \frac{\dot{M}_{\ast}(t)}{\sum\limits_j{J_j}}
    \end{equation}
    where \(\dot{M}_{\ast}(t)\) is the stellar escape rate from \clusterBH, and
    \(J_j\) is the integration over \(mf(m)h(m)\) in each mass bin \(j\), such
    that the mass loss rate in each bin is \(\dot{M}_j=-C J_j\).
    In \MMpapers, this unknown quantity was simply set to \(\dot{M}_{\ast}=0\),
    meaning that the inferred IMF slopes effectively
    represented the PDMF, rather than the true IMF.
    In this work, however, this extra step in the
    modelling means we are able to directly infer the actual IMF slopes.

\subsection{Fitting of \coupled Models}


    In this work, we use the models described in \Cref{sub:coupled_models}
    to fit the observational data available for a large sample of MW
    GCs and explore their initial conditions.

    \subsubsection{Cluster Data}
    \label{ssub:cluster_data}

    For this analysis we return to the same sample of clusters chosen in
    \MMpapers, which were selected to provide a large sample of MW GCs
    with sufficient quantity and quality of kinematic and stellar mass function
    data.

    While this sample of 40 MW clusters allows us to survey the
    conditions of a significant population of the MW GCs, it should be noted,
    as discussed in \MMpaperII, that the clusters chosen are, by virtue of the
    required data quality, biased slightly towards more nearby, massive,
    and denser clusters.

    The observational datasets used for each cluster are also the same as those
    used in \MMpapers (with a few small additions), and are briefly
    summarised below.

    Binned radial profiles of proper motion (PM) dispersion, in both the radial
    and tangential directions, are provided by \Gaia
    (\citealt{GaiaCollaboration2023}; \citetalias{Dickson2023})
    in the cluster outskirts and by the Hubble Space
    Telescope \citep[\HST;][]{Watkins2015,Libralato2022} in the cores of most
    clusters in our sample. We also include the new PM dispersion profiles of
    \omegacen from \citet{Haberle2025}.
    Line-of-sight (LOS) velocity dispersion profiles are taken from
    compilations of various ground-based
    \citep{Lutzgendorf2013,Baumgardt2018,Kamann2018,Dalgleish2020} and \Gaia
    \citep{Baumgardt2019a} datasets.
    Projected number density profiles for most clusters are taken from
    \citet{deBoer2019}, who combined \Gaia star counts in the outskirts
    and \HST counts \citep{Miocchi2013} or ground-based surface-brightness
    profiles \citep[SBPs,][]{Trager1995} in the central regions.
    Three of our clusters (\NGC4372, \NGC4833 and \NGC5927) do not appear in
    this catalogue, and instead the surface brightness profiles compiled by
    \citet{Baumgardt2017}, based on data from
    \citet{Melbourne2000,Sarajedini2007,Kacharov2014} are used
    \footnote{These three clusters were excluded from \MMpaperI due to the lack
    of number density data, but are now included in this study. All other
    clusters in our sample were present in \MMpaperI.}.
    Finally, present-day local stellar mass functions are taken from
    \citet{Baumgardt2023}, and consist of star counts extracted from a large
    number of \HST pointings in each cluster.

    The metallicities of the clusters are taken from \citet{Harris1996}.
    The ages of most clusters are taken from the catalogue of
    \citet{Vandenberg2013}. Four clusters not in this catalogue instead have
    their ages taken from \citet[][\NGC4372, \NGC6266]{DeAngeli2005} and
    \citet[][\NGC5139, \NGC6093]{MarinFranch2009}.
    However, our results are relatively insensitive to the exact ages
    chosen, except in cases where the clusters are actively dissolving, which
    do not appear in our chosen sample.

    As discussed in \Cref{sub:evolutionary_models}, each cluster must also be
    placed on a circular orbit in \clusterBH. The galactocentric radius of this
    orbit is held constant throughout the evolution; however, in reality,
    the true orbit of clusters may be notably eccentric. In order to account
    for this, we follow the prescription described by
    \citet{Baumgardt2003,Alexander2014}
    to compute an ``effective circular orbit'' at the galactocentric radius:
    \begin{equation}
        \RGeff = R_{\mathrm{a}} (1-e) = \frac{2 R_{\mathrm{a}} R_{\mathrm{p}}}{R_{\mathrm{a}} + R_{\mathrm{p}}}
    \end{equation}
    where \(R_{\mathrm{a}}\) and \(R_{\mathrm{p}}\) are the apocentre and
    pericentre of a cluster's orbit, taken from \citet{Vasiliev2021}, and \(e\)
    is the orbital eccentricity.
    Placing clusters on this circular orbit roughly matches the tidal
    effects of the true eccentric orbits over the lifetime of the cluster
    \citep{Baumgardt2003,Cai2016}.

    \subsubsection{Fitting Procedure}
    \label{ssub:fitting}

    In order to infer the posterior probability distributions of the model
    parameters that best describe each cluster, we utilize the same Bayesian
    parameter estimation methods as in \MMpaperI. In particular,
    the \coupled models are constrained by the datasets described above and
    the posterior is sampled using dynamic nested sampling (by
    the dynesty software package \citealp{Speagle2020}), through the \GCfit
    fitting pipeline.
    The free parameters inferred in this work are listed
    in \Cref{tab:free_params} along with the adopted priors. In the same
    fashion as \MMpaperI, we adopt, for most parameters, a uniform prior with
    limits chosen to bound a wide enough range of parameter space to be
    largely uninformative, or to cover the valid physical range of the
    parameter.
    For the heliocentric distance \(d\), we instead adopt a Gaussian prior,
    with a mean and width taken from the averaged distances computed by
    \citet{Baumgardt2021}.

    Finally, to place some constraints on the expected relationship
    between the cluster core sizes and BH subsystems, which may not be
    discernible in the given observational data alone, we also include
    a regularization prior on the size of the cluster cores (\rc).
    By examining the full grid of CMC models, a clear relationship emerges
    between \rc\ and the total mass in BHs \citep{Kremer2025}.
    We thus fit a linear relation between \fbh and
    \(\log\left(\rc/\rh\right)\) (using the ``theoretical'' core
    radius from \citealt{Casertano1985}) in the CMC models, finding a slope,
    intercept and scatter of approximately 0.5, -0.5 and 0.2 respectively, and
    include a Gaussian, informative prior probability term centred on this
    relation to the posterior.
    However, tests without this extra constraint reveal that it does not
    typically have a large impact on the results presented here, with only a
    handful of core-collapsed clusters (discussed in \Cref{sub:outliers})
    straying from the expected relation.

    \begin{table}
    \renewcommand*{\arraystretch}{1.8}
    \centering
    \begin{tabular}{ c c c }
        \hline
        Parameter                      & Description & Prior \\
        \hline
        \(\hat{\phi}_0\)               & \makecell[l]{Dimensionless \\ central potential}                        & \(\mathcal{U}(0.0,\, 100.0)\) \\
        \(g\)                          & \makecell[l]{Truncation \\ parameter}                                   & \(\mathcal{U}(0.0,\, 3.5)^\ast\) \\
        \(\log(\hat{r}_{\mathrm{a}})\) & \makecell[l]{Dimensionless \\ anisotropy radius}                        & \(\mathcal{U}(-2.0,\, 8.0)\) \\
        \(\delta\)                     & \makecell[l]{Velocity-scale \\ mass dependence}                         & \(\mathcal{U}(0.1,\, 0.5)\) \\
        \(\zeta\)                      & \makecell[l]{Velocity-scale \\ high-mass scaling}                       & \(\mathcal{U}(0.01,\, 1.0)\) \\
        \(\eta\)                       & \makecell[l]{Anisotropy-scale \\ mass dependence}                       & \(\mathcal{U}(-5.0,\, 5.0)\) \\
        \hline
        \(\Mi\)                        & \makecell[l]{Initial total cluster \\ mass \(\left[10^6\ \Msun\right]\)} & \(\mathcal{U}(0.01,\, 5.0)\) \\
        \(\rhi\)                       & \makecell[l]{Initial half-mass \\ radius \([\mathrm{pc}]\)}             & \(\mathcal{U}(0.01,\, 10.0)\) \\
        \(\alpha_1\)                   & \makecell[l]{IMF exponent \\ \((0.1\ \Msun<m\leq 0.5\ \Msun)\)}         & \(\mathcal{U}(-1.0,\, 4.0)^\ast\) \\
        \(\alpha_2\)                   & \makecell[l]{IMF exponent \\ \((0.5\ \Msun<m\leq 1\ \Msun)\)}           & \(\mathcal{U}(-1.0,\, \alpha_1)^\ast\) \\
        \hline
        \(F\)                          & \makecell[l]{Mass function \\ nuisance parameter}                       & \(\mathcal{U}(1.0,\, 5.0)\) \\
        \(s^2\)                        & \makecell[l]{Number density \\ nuisance parameter}                      & \(\mathcal{U}(0.0,\, 15.0)\) \\
        \(d\)                          & \makecell[l]{Heliocentric \\ distance \([\mathrm{kpc}]\)}               & \(\mathcal{N}(d_{\mathrm{lit}},\, \delta d_{\mathrm{lit}})\) \\
        \hline
    \end{tabular}
    \caption{
        List of all free parameters, their descriptions and the prior
        probability distributions adopted here.
        The first six are structural \limepy parameters, the next four
        define the initial cluster conditions, and the final three parameters
        aid in comparing models to observations.
        Uniform priors (\(\mathcal{U}\)) are indicated alongside their upper
        and lower bounds, while the prior on the distance is a Gaussian
        distribution (\(\mathcal{N}\)) with mean \(d_{\mathrm{lit}}\) and
        width \(\delta d_{\mathrm{lit}}\) taken from \citet{Baumgardt2021}.
        Prior bounds motivated by physical or model constraints are denoted
        by an asterisk. All others are chosen to bound a large enough area of
        parameter space to be largely uninformative.
        We note that the upper bound on the \Mi prior is increased to
        \(20\times10^6\,\Msun\) for the massive clusters \NGC{104} and
        \NGC{5139}.
    }
    \label{tab:free_params}
    \end{table}


\section{Validation}\label{sec:validation}

    To test the ability of our \coupled
    models to reproduce the evolution of realistic clusters and recover their
    initial conditions, we first apply this fitting method to simulated
    present-day observations of star-by-star dynamical models, namely the
    existing grid of CMC models presented in
    \citet{Kremer2020}\footnote{Available at
    \url{https://cmc.ciera.northwestern.edu}}.

    Simulations are selected from the available catalogue to cover
    a wide range of realistic MW GC conditions. To this end, we
    select all low-metallicity (\(Z=0.0002,\ 0.002\)) models in the catalogue
    (except for the smallest; \(N_0=[2, 4]\times 10^5\)), as well as eight
    of the higher-metallicity clusters (\(Z=0.02\); spanning the available
    initial radii for \(N_0=[8, 16, 32]\times 10^5\) models at
    \(R_{\mathrm{G}}=\SI{8}{\kilo\pc}\)).
    Five of these selected models were subsequently excluded from our sample
    due to being either tidally disrupted before the end of the simulation,
    or being terminated early due to ``collisional runaway'' effects
    (see \citealt{Kremer2020} for details).
    This yields a final sample of 53 CMC models. For each model we
    choose the final snapshot in the catalogue to represent the
    present-day conditions of the mock clusters, resulting in
    a final age of around 13 Gyr for all clusters.

    These models, which have circular galactic orbits with galactocentric
    distances of \(R_{\mathrm{G}}=20\), 8 and \SI{2}{\kilo\pc}, are placed at
    heliocentric distances of 8, 6.1 and \SI{5}{\kilo\pc}, typical of the
    distances seen in our sample of MW clusters.
    We then utilize the \textsc{cmctoolkit} library \citep{Rui2021b,
    Rui2021a} to compute projected positions, velocities and simulated
    photometry for stars in each snapshot. From these, we extract synthetic
    observations from each selected snapshot, designed to emulate the
    real observational datasets described in
    \Cref{ssub:cluster_data}.
    For more details on the generation of these mock observations, see
    section 3.1 of \MMpaperII.

    We then fit the \coupled models to these mock observations using the same
    priors as for the real clusters
    The width of the distance prior is set to
    \(\sigma=\SI{0.1}{\kilo\pc}\), to mimic realistic uncertainties.
    The fitting proceeds exactly as done when fitting on real
    clusters, with no information about the true evolutionary history of the
    models provided.

\subsection{Results of Fits to Mocks}
\label{sub:results_of_fits_to_mocks}

\begin{figure}
        \centering
        \includegraphics[width=\linewidth]{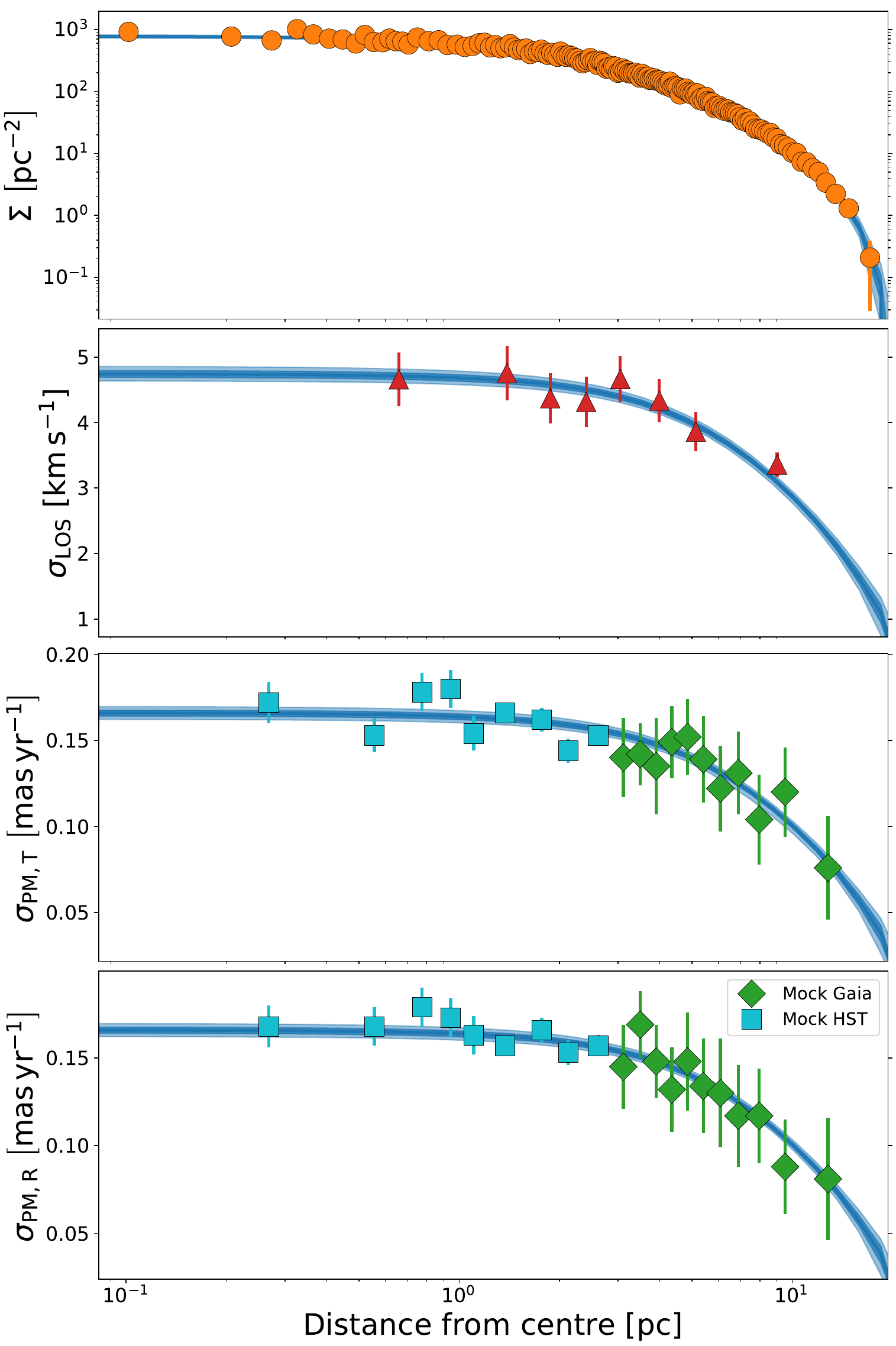}
        \caption{
            Model radial profiles (blue contours) of
            surface number density (\(\Sigma\)),
            line-of-sight velocity dispersions (\(\sigma_{\mathrm{LOS}}\)),
            radial (\(\sigma_{\mathrm{PM},\mathrm{R}}\)) and
            tangential (\(\sigma_{\mathrm{PM},\mathrm{T}}\)) proper motion
            dispersions, for the fit of the mock observations of
            the CMC simulation with initial conditions
            \(N_0=8\times10^5\), \(r_{\mathrm{v},0}=\SI{2}{\pc}\),
            \(R_{\mathrm{G}}=\SI{2}{\kilo\pc}\) and \(Z=0.0002\).
            The dark and light shaded regions represent the \(1\sigma\)
            and \(2\sigma\) credible intervals of the model fits, respectively.
            The mock observational datasets used to constrain the models are
            shown alongside their\(1\sigma\) uncertainties by the various
            markers and errorbars.
        }
        \label{fig:mock_profile_fit}
    \end{figure}

\begin{figure*}
        \centering
        \includegraphics[width=\linewidth]{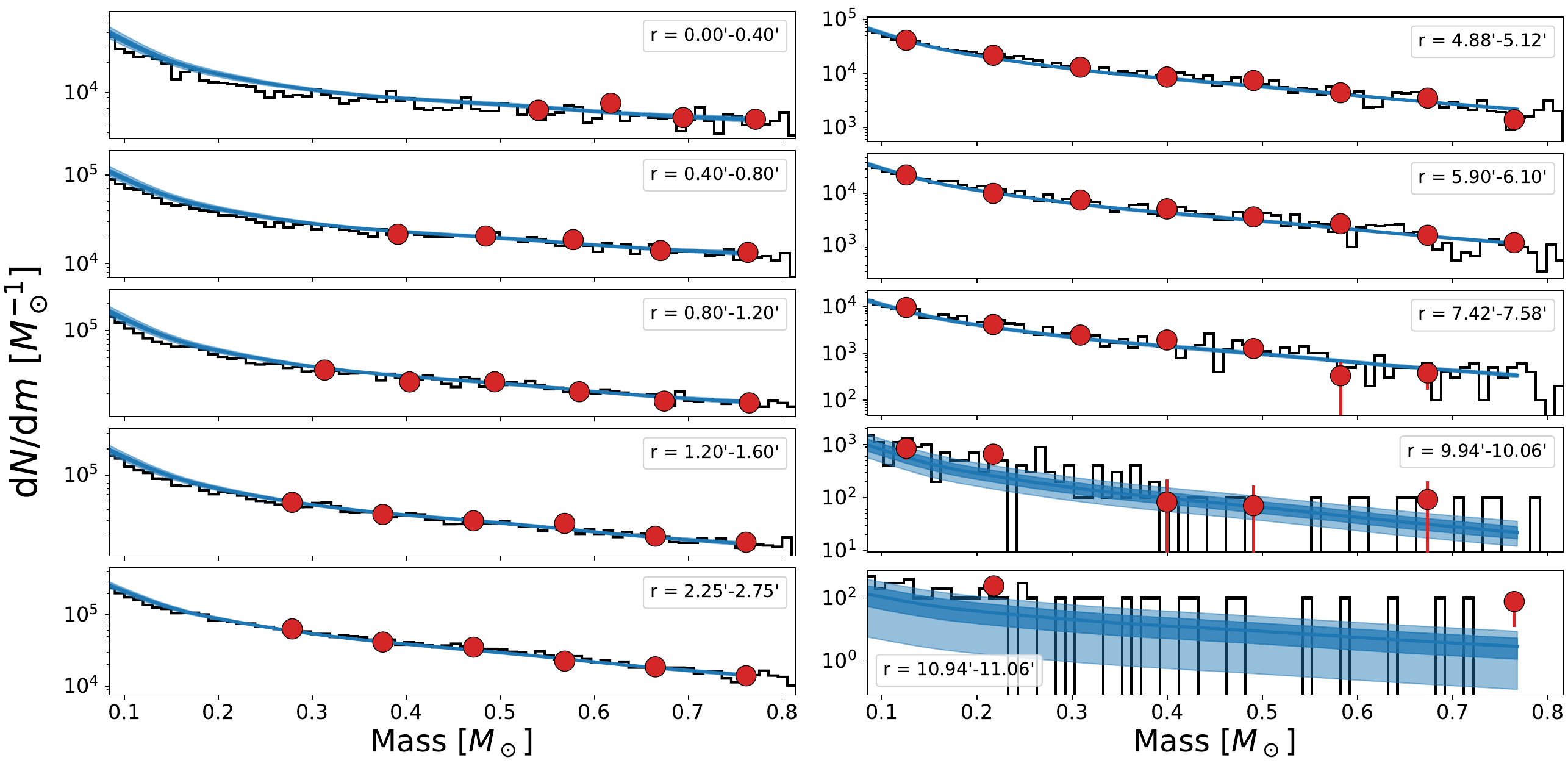}
        \caption{
            Model present-day local stellar mass functions (blue contours)
            for the fit of the mock observations of
            the CMC simulation with initial conditions
            \(N_0=8\times10^5\), \(r_{\mathrm{v},0}=\SI{2}{\pc}\),
            \(R_{\mathrm{G}}=\SI{2}{\kilo\pc}\) and \(Z=0.0002\).
            Each panel shows the number of stars per unit mass
            as a function of stellar mass, for different projected distance
            ranges from the cluster centre.
            The dark and light shaded regions represent the \(1\sigma\)
            and \(2\sigma\) credible intervals of the model fits, respectively.
            The mock observational datasets used to constrain the models are
            shown alongside their \(1\sigma\) uncertainties by the red
            circles and errorbars.
            The true underlying mass function from the full CMC snapshot is
            shown in black.
        }
        \label{fig:mock_massfunc_fit}
    \end{figure*}

\begin{figure}
        \centering
        \includegraphics[width=\linewidth]{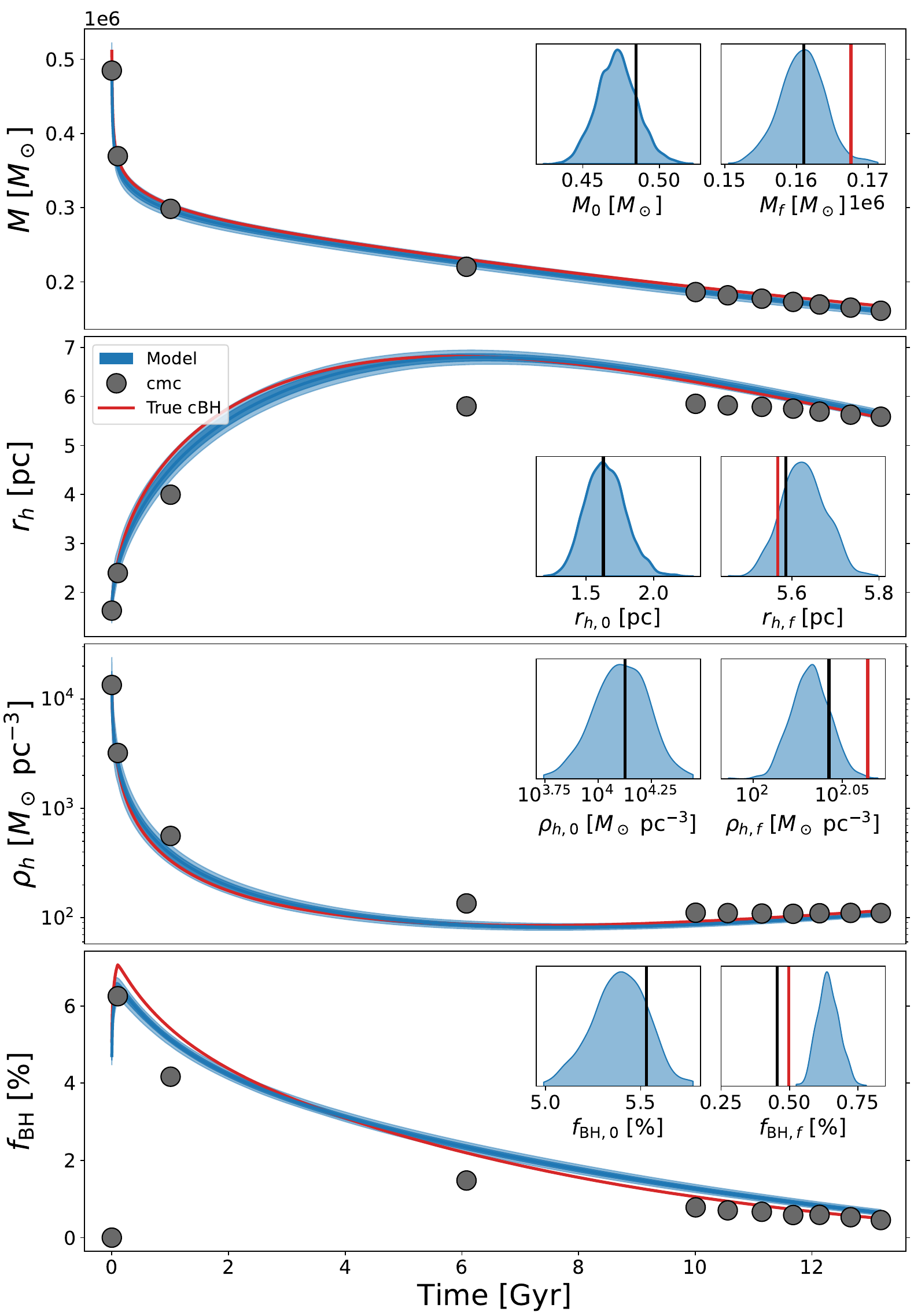}
        \caption{
            Evolution of the model total mass, half-mass radius, half-mass
            density and BH mass fraction over time (blue contours) for the fit
            of the mock observations of the CMC simulation with initial
            conditions \(N_0=8\times10^5\), \(r_{\mathrm{v},0}=\SI{2}{\pc}\),
            \(R_{\mathrm{G}}=\SI{2}{\kilo\pc}\) and \(Z=0.0002\).
            The dark and light shaded regions represent the \(1\sigma\)
            and \(2\sigma\) credible intervals of the model fits,
            respectively.
            The true values from the CMC snapshots are shown as black
            circles. Note that the models are not fit on these quantities,
            but they serve to illustrate the recovery of
            the evolutionary history of the cluster
            by fitting to the present-day (mock) observations alone.
            Shown in red is the evolution of a single \clusterBH model when
            initialized with the true initial conditions of
            the corresponding CMC model.
            The inset panels show the posterior distributions from our fit
            of the initial and present-day values of each quantity in blue,
            with the true values from the CMC models indicated by the black
            vertical lines.
        }
        \label{fig:mock_evolution_fit}
    \end{figure}

\begin{figure*}
        \centering
        \includegraphics[width=\linewidth]{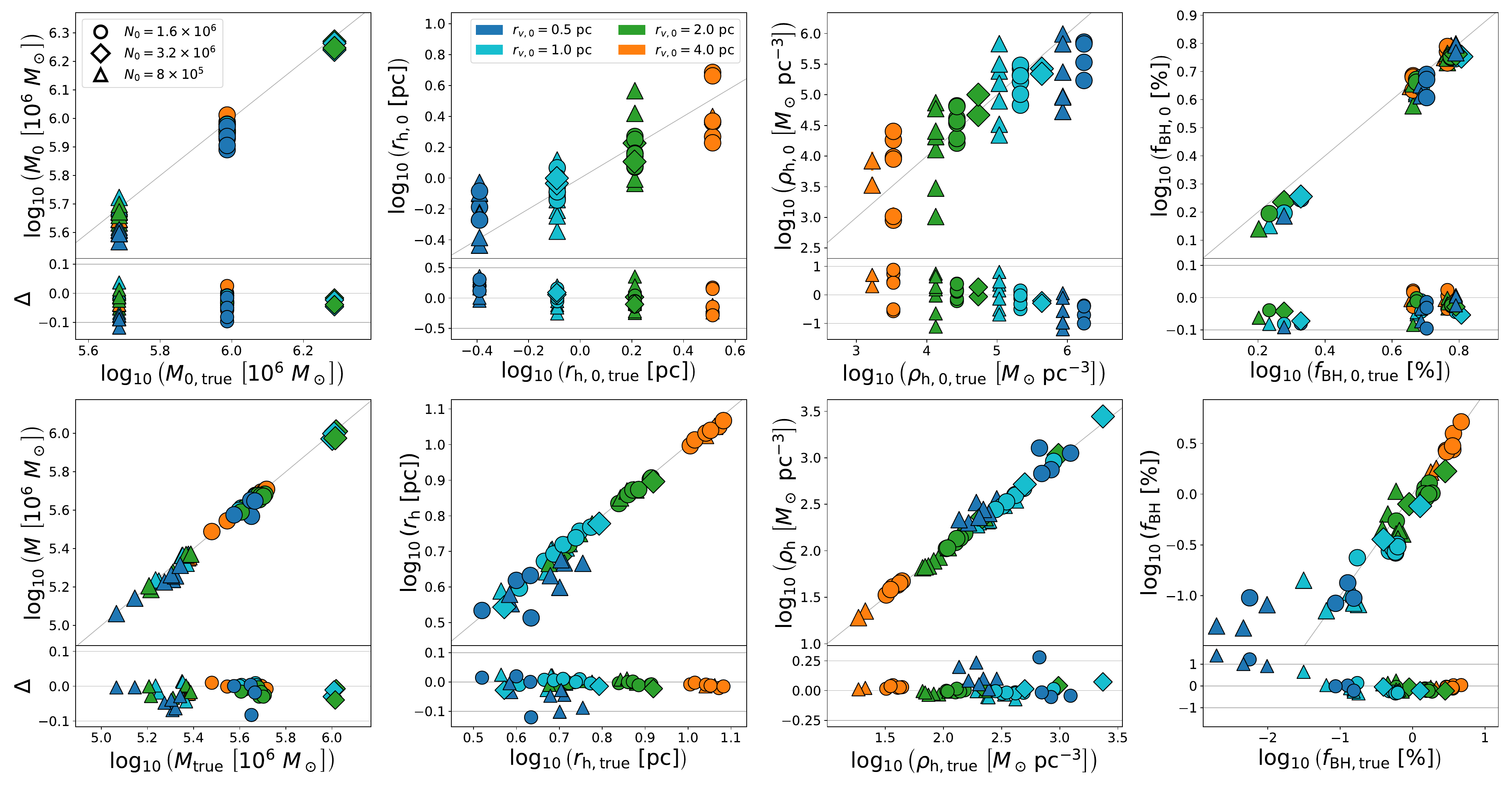}
        \caption{
            Comparison of the initial (top row) and present-day (bottom row)
            values of the total mass, half-mass radius, half-mass density and
            mass fraction in BHs inferred by our best-fitting models
            (on the y-axis),
            against the true values from the CMC models (on the x-axis).
            The points are coloured based on their true initial radii, and shown
            by a diamond, circle or triangle based on their initial number of
            stars (\(N_0=[32, 16, 8]\times 10^5\), respectively).
            Note that, due to the coarse snapshot time resolution
            available in the public CMC grid, the \fbhi values are not taken
            directly from the CMC snapshots but are instead recomputed based on
            the true initial CMC conditions using the same \ssptools methods
            as in our models.
            This allows for the most direct comparison between the inferred
            and true values.
            The residuals (inferred minus true, in dex) are shown as a
            function of the true values below each panel.
        }
        \label{fig:mock_param_comp}
    \end{figure*}

    \Cref{fig:mock_profile_fit,fig:mock_massfunc_fit} show an example of the
    resulting fits of our models to these mock observational datasets, for
    the CMC simulation with initial conditions \(N_0=8\times10^5\),
    \(r_{\mathrm{v},0}=\SI{2}{\pc}\), \(R_{\mathrm{G}}=\SI{2}{\kilo\pc}\),
    \(Z=0.0002\) (corresponding to model number 75 in \citealt{Kremer2020}).
    \Cref{fig:mock_evolution_fit} shows the recovery of the evolution over time
    of the bulk quantities of this same simulation.
    In \Cref{fig:mock_param_comp} we show the comparison between the true CMC
    values and our inferred values, for both the initial and present-day
    mass, radius, density and BH mass fraction, for the full sample of mocks
    we have examined.
    Our model fitting is able to successfully reproduce all of the simulated
    present-day kinematic, number density and mass function datasets, for
    all of the CMC models in our sample\footnote{Similar figures showing the
    fits to all models in the sample, as well as full sampler chains, can
    be found at http://doi.org/10.11570/26.0014.}.
    When comparing the inferred evolution of our models over time against the
    available snapshots of the CMC models, we see that despite fitting
    only to the present-day observables, we generally recover the
    evolution and initial conditions very well in most clusters.

    It is clear from these figures that we are able to recover the present-day
    conditions of the entire mock cluster sample very well, where these
    quantities are directly constrained by the observational data.
    At most, we recover the mass, radius, density and BH mass fraction to
    within \(10^5\,\Msun\), \SI{1}{\pc}, \SI{0.25}{\dex} and \SI{1}{\%}
    respectively, with the majority of clusters performing better than this.
    A handful of models with very low present-day values of \fbh
    are overestimated more significantly, however these
    correspond to models with almost no BHs remaining to the present-day,
    and the differences in \limepy models with between \(\fbh=0\) and
    \(<0.1\%\) are nearly indistinguishable (the differences are amplified
    in \Cref{fig:mock_param_comp} by the log-scale).

    The initial conditions are also recovered well, however
    there is a notably larger spread in the inferred quantities.
    At most, we recover the initial values of mass, radius, density and
    BH mass fraction to within \(2.5\times10^5\,\Msun\), \SI{2}{\pc},
    \SI{1}{\dex} and \SI{1}{\%} respectively, with most clusters again
    performing more accurately than this.
    In particular, there is a spread among the initial densities, driven largely
    by smaller discrepancies in the initial radii (which are amplified as
    \(\rhoh\propto\rh^{-3}\)), though no strong systematic bias towards higher
    or lower densities is evident here.

    It is likely that most issues in inferring the radii appear due to
    discrepancies in the \clusterBH models themselves, when compared to CMC.
    Due to the fact that our models are constrained only by the present-day
    observables,
    any models for which \clusterBH (when starting with the true
    initial conditions) under(over)-estimates the final conditions will likely
    cause our inferred initial conditions to be correspondingly
    over(under)-estimated, to ensure the correct present-day conditions
    are recovered.
    For example, if \clusterBH slightly underestimates the cluster expansion,
    recovering the correct present-day radius will require a slight
    overestimation of the initial radius, to compensate for this discrepancy.
    In \Cref{fig:mock_evolution_fit}, the evolution of \clusterBH models
    beginning with the true initial conditions is shown in red,
    to illustrate this effect.
    In \cbhpaper, the \clusterBH models were typically found to be in agreement
    with the CMC models to within 20-30 per cent in total mass, half-mass
    radius and BH mass over their entire evolution, and thus we might
    expect, at a minimum, corresponding residuals in our recovery of the initial
    quantities. The agreement between \clusterBH and CMC also varies across
    the grid, with some regions of parameter space
    (e.g. \(r_{\mathrm{v},0}=\SI{4}{\pc}\)) expected to exhibit larger residuals
    than others.

    As was found in Section 3.2 of \MMpaperII,
    when fitting multimass DF-based models to the same observables,
    the statistical uncertainties derived from the fitting
    noticeably underestimate the overall true uncertainties required in
    quantities such as the present-day BH mass fraction and the initial mass and
    radius.
    As was originally quantified in \MMpaperII,
    this underestimation may arise because our procedure operates under the
    assumption that our models are able to perfectly represent the data, and
    thus may underestimate the true errors if there is any misspecification
    between the models we assume and the true underlying
    models from which the data is generated.
    Thus we expect that there remain unquantified systematic errors beyond the
    statistical uncertainties reported here.

    Despite these caveats, this validation against mock observations
    demonstrates that the coupled \coupled models provide us with a useful
    tool, which can reliably predict the initial conditions of GCs based
    solely on their present-day, observable properties, across a large
    volume of parameter space.




\section{Fitting to Milky Way Clusters}\label{sec:real_fitting}


    In this section we present the results of applying our fitting procedure,
    as described in \Cref{ssub:fitting}, to our full
    sample of Milky Way GCs, as described in \Cref{ssub:cluster_data}.

\begin{figure}
        \centering
        \includegraphics[width=\linewidth]{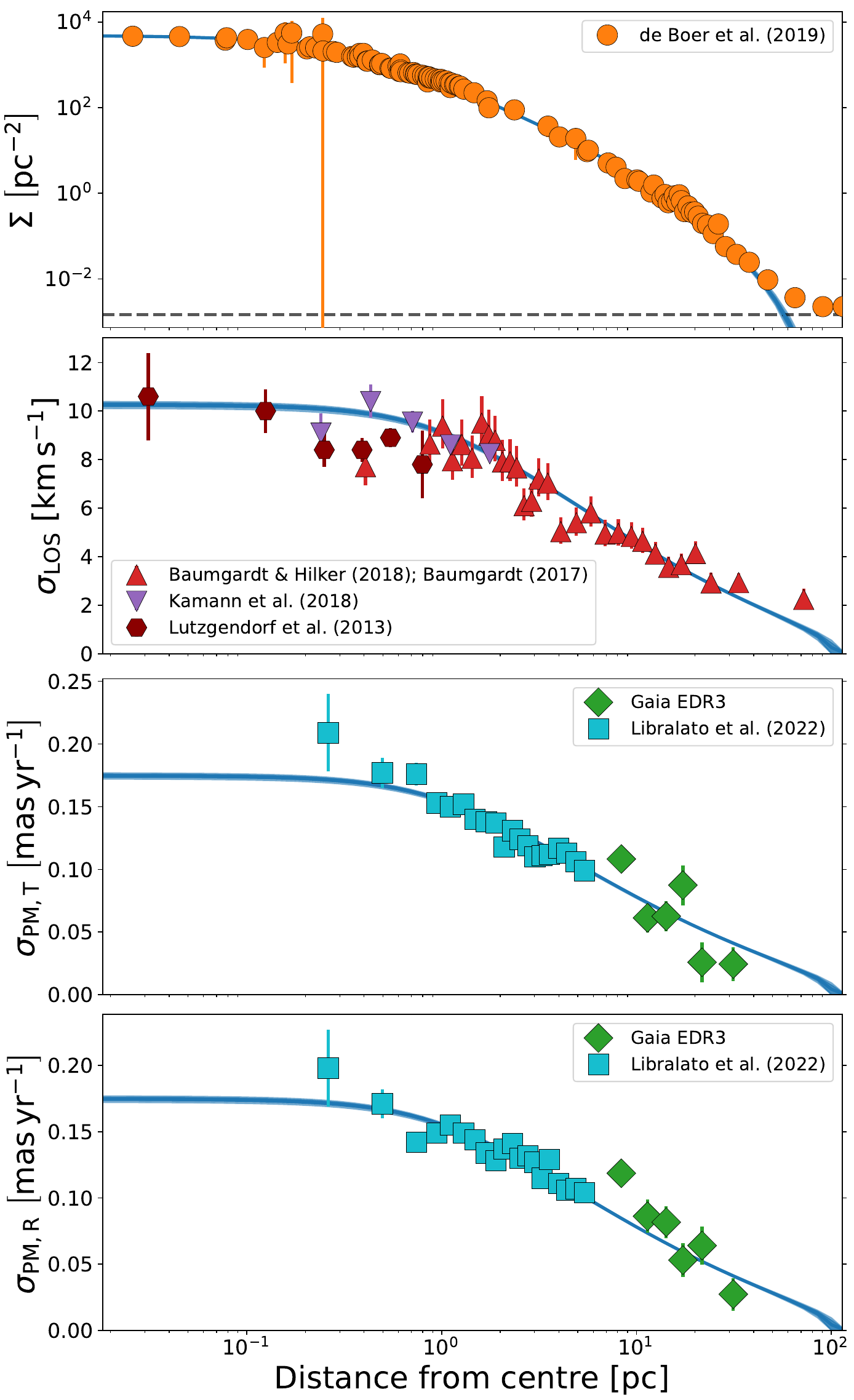}
        \caption{
            Model radial profiles (blue contours) of
            surface number density (\(\Sigma\)),
            line-of-sight velocity dispersions (\(\sigma_{\mathrm{LOS}}\)),
            radial (\(\sigma_{\mathrm{PM},\mathrm{R}}\)) and
            tangential (\(\sigma_{\mathrm{PM},\mathrm{T}}\)) proper motion
            dispersions, for the fit of \NGC{1851}.
            The dark and light shaded regions represent the \(1\sigma\)
            and \(2\sigma\) credible intervals of the model fits, respectively.
            The observational datasets used to constrain the models are shown
            alongside their \(1\sigma\) uncertainties by the various
            markers and errorbars.
            The background level subtracted from the number density profile is
            shown by the dashed line.
        }
        \label{fig:real_profile_fit}
    \end{figure}

\begin{figure*}
        \centering
        \includegraphics[width=\linewidth]{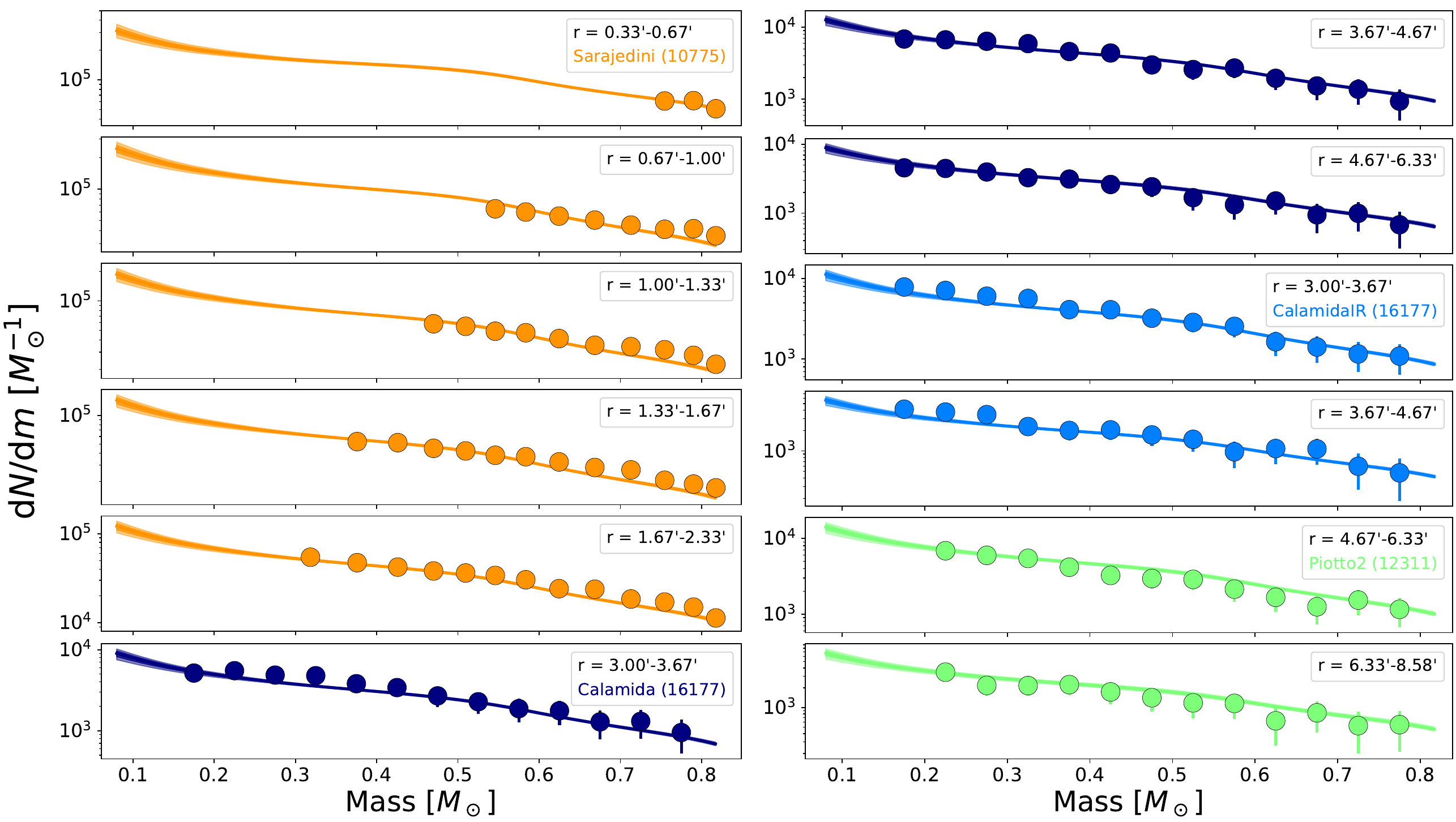}
        \caption{
            Model present-day local stellar mass functions for the fit of
            \NGC{1851}.
            Each panel shows the number of stars per unit mass
            as a function of stellar mass, for different projected distance
            ranges from the cluster centre.
            The dark and light shaded regions represent the \(1\sigma\)
            and \(2\sigma\) credible intervals of the model fits, respectively.
            The measurements used to constrain the models are shown alongside
            their \(1\sigma\) uncertainties by the circles and errorbars.
            The panels are colour-coded by the \textit{HST} program used, with
            the PI name and identifier listed in the top-right
            corners. Note that some programs overlap in their radial coverage.
        }
        \label{fig:real_massfunc_fit}
    \end{figure*}

\begin{figure}
        \centering
        \includegraphics[width=\linewidth]{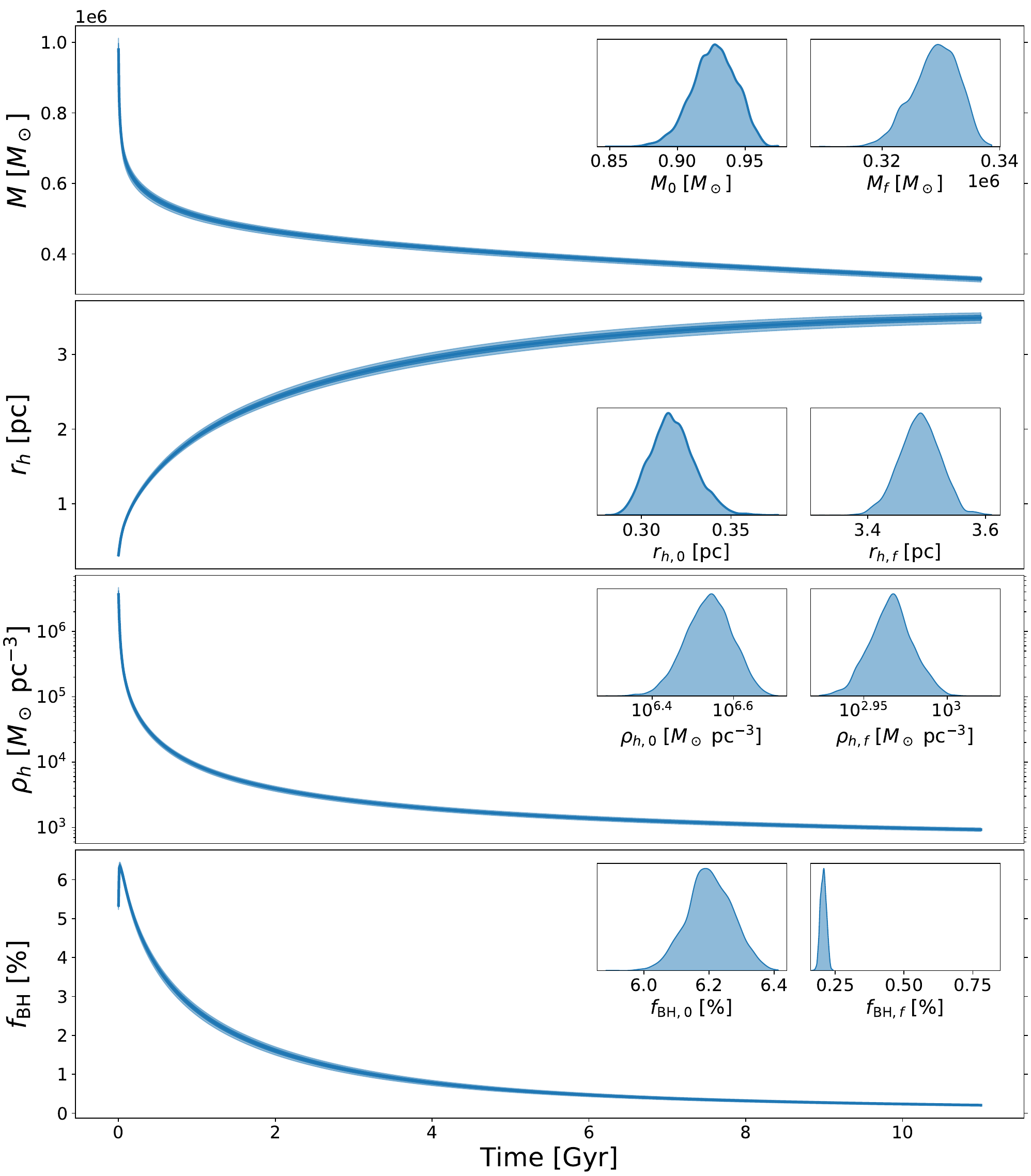}
        \caption{
            Evolution of cluster total mass, half-mass radius, half-mass density
            and BH mass fraction over time for the fit of \NGC{1851}.
            The dark and light shaded regions represent the \(1\sigma\)
            and \(2\sigma\) credible intervals of the model fits,
            respectively.
            The inset panels show the posterior distributions from our fit
            of the initial and present-day values of each quantity.
        }
        \label{fig:real_evolution_fit}
    \end{figure}

    In \Cref{fig:real_profile_fit} and \Cref{fig:real_massfunc_fit}, we show
    an example fit of our models to the observational datasets
    of the cluster \NGC{1851}, while in \Cref{fig:real_evolution_fit} we show
    the inferred evolution of the mass, radius, density and \(\fbh\) from
    initial conditions to their present-day values.
    The best-fitting models provide good fits to the data in the majority of
    clusters in our sample\footnote{Similar figures showing the fits
    to all models in the sample, as well as full sampler chains, can
    be found at http://doi.org/10.11570/26.0014.}.
    The best-fitting parameter values for all clusters can be found in
    \Cref{table:best_fitting_params}, while derived quantities such as the
    half-mass densities, BH mass fractions, and present-day total mass and
    half-mass radius can be found in \Cref{table:extra_params}.

\subsection{Outliers and Unsatisfactory Fits}
\label{sub:outliers}

    There are a small number of clusters in our MW sample to which the
    fits do not reproduce certain observational datasets adequately, or that
    exhibit other problematic behaviours, which we either flag or remove
    from our sample entirely. These fits are discussed below.


    In certain clusters the stellar mass function data, which consists of
    various HST pointings divided into radial slices, include multiple fields
    which overlap with one another in projected distance from the cluster
    centre.
    In these cases, the local mass functions are not
    always entirely consistent with each other within the same radial bin.
    This may reflect inconsistencies in the stellar photometry or isochrone
    fitting of \citet{Baumgardt2023}. In most of these cases, such as
    in \NGC{5904}, \NGC{6752} or \NGC{7099}, the differences are small enough
    that they do not seriously affect the fits, given our mass function nuisance
    parameter \(F\) accounting for such systematic uncertainties.
    However, in \NGC{6656}, these discrepancies are severe enough
    that the fits to the data are unsatisfactory, and we therefore remove this
    cluster from our sample.

    A few clusters in our sample do not adequately reproduce
    the kinematic data (and thus the dark mass). This often occurs due to
    a lack of data in some dimension (e.g. PM or LOS dispersion profiles),
    which may lead to the overfitting of other datasets such as the mass
    function or density profiles.
    We therefore also remove \NGC{288}, \NGC{5927} and \NGC{6981} from our
    sample.


    Several clusters in our sample are also classified as
    ``core-collapsed'' by, e.g., \citet{Trager1995}.
    As was described in more detail in \MMpaperII, visibly core-collapsed
    clusters are expected to harbour few, if any, BHs today as, in the presence
    of a BH subsystem, the efficient heat transfer from BHs to stars causes
    the visible core to remain large relative to \rh, preventing the
    collapse of the visible stars until nearly all BHs have been ejected
    \citep{Giersz2009,Breen2013,Chatterjee2013,Kremer2020b}.
    However, as shown in \Cref{sub:black_hole_populations}, for some of our
    core-collapsed clusters (e.g. \NGC{7078}) we do infer a significant number
    of BHs.
    Core-collapsed clusters have cuspy inner surface brightness profiles,
    which may be difficult to reproduce with \limepy models, as they are
    cored by definition. These clusters also stand out as the only ones which
    stray notably from the theoretical \(\rc - \fbh\) relationship.
    In \MMpaperII we conducted a separate analysis with these clusters by fixing
    their BH retention to 0; however, this would be difficult to recreate under
    the present methodology, as the BH retention is tied to the evolution of
    the other model parameters, rather than set independently.
    Therefore, the fits to these clusters, especially those inferring
    significant BH populations, should be regarded with caution.
    All core-collapsed clusters are denoted by an asterisk in the rest of this
    manuscript.


    Finally, there is one model assumption which is not captured in the
    fits to mock data which could notably affect our fits
    to the real clusters: the cluster orbits.
    While we place our GCs on circularized orbits within a simple isothermal
    sphere potential, in reality over their entire lifetimes the orbits of
    clusters may see them interact with galactic substructures, sink towards the
    galactic centre via dynamical friction, be accreted from ex-situ sources,
    or pass through the galactic disc.
    Likewise, the orbital parameters we take from
    \citet{Vasiliev2021} have associated uncertainties, which would only
    increase when extrapolated further back in time, especially as the potential
    of the galaxy evolves itself.
    For most clusters (excluding significant perturbations by structures
    about which we have no knowledge), where the tides do not dominate the
    mass loss, small changes in the effective galactocentric radius do not have
    strong impacts on the evolution or the parameters we infer.
    However, the impacts of slight changes in \RGeff increase as \RGeff
    decreases.
    Some clusters with notably small effective galactocentric radii
    (\(\RGeff\lesssim\SI{1.5}{\kilo\pc}\); well below
    the range covered by the CMC grid used to validate our models)
    are thus particularly sensitive to uncertainties in their orbital
    parameters and history. For example, testing demonstrates that an increase
    of the \RGeff of \NGC{6121} of only \SI{0.2}{\kilo\pc} (up to
    \(\RGeff\sim\SI{1.3}{\kilo\pc}\)) can result in an increase in the inferred
    initial density of around \SI{0.5}{\dex}.
    This is further supported by comparison with Monte Carlo modelling of this
    cluster from \citet{Heggie2008}, where it was placed at a higher orbit,
    and yielded a significantly more compact initial size
    (\(\rhi=\SI{0.58}{\pc}\), compared to our \(\rhi=\SI{1.7}{\pc}\)).
    At these small galactocentric distances, the discrepancies between our
    chosen model of the galactic potential (which has no defined core) and the
    true MW potential, may also become more apparent.
    Given this, the fits to these clusters should therefore be regarded with
    increased caution.
    All clusters with notably small \RGeff are denoted by a dagger in the rest
    of this paper.
    This especially includes \NGC{6093}, \NGC{6121} and \NGC{6266}, which are
    all prominent outliers in our inferred initial density distribution
    (see \Cref{sub:initial_conditions}).
    \NGC{6624}, the cluster in our sample with the smallest \RGeff
    (\SI{0.5}{\kilo\pc}), was unable to be fit by the models at all, and is
    removed from our sample entirely.

    This leaves us with a final cluster sample of 35 MW GCs, which are
    used in all subsequent analysis.

\subsection{Comparison with Validation Sample}

\begin{figure*}
        \centering
        \includegraphics[width=\linewidth]{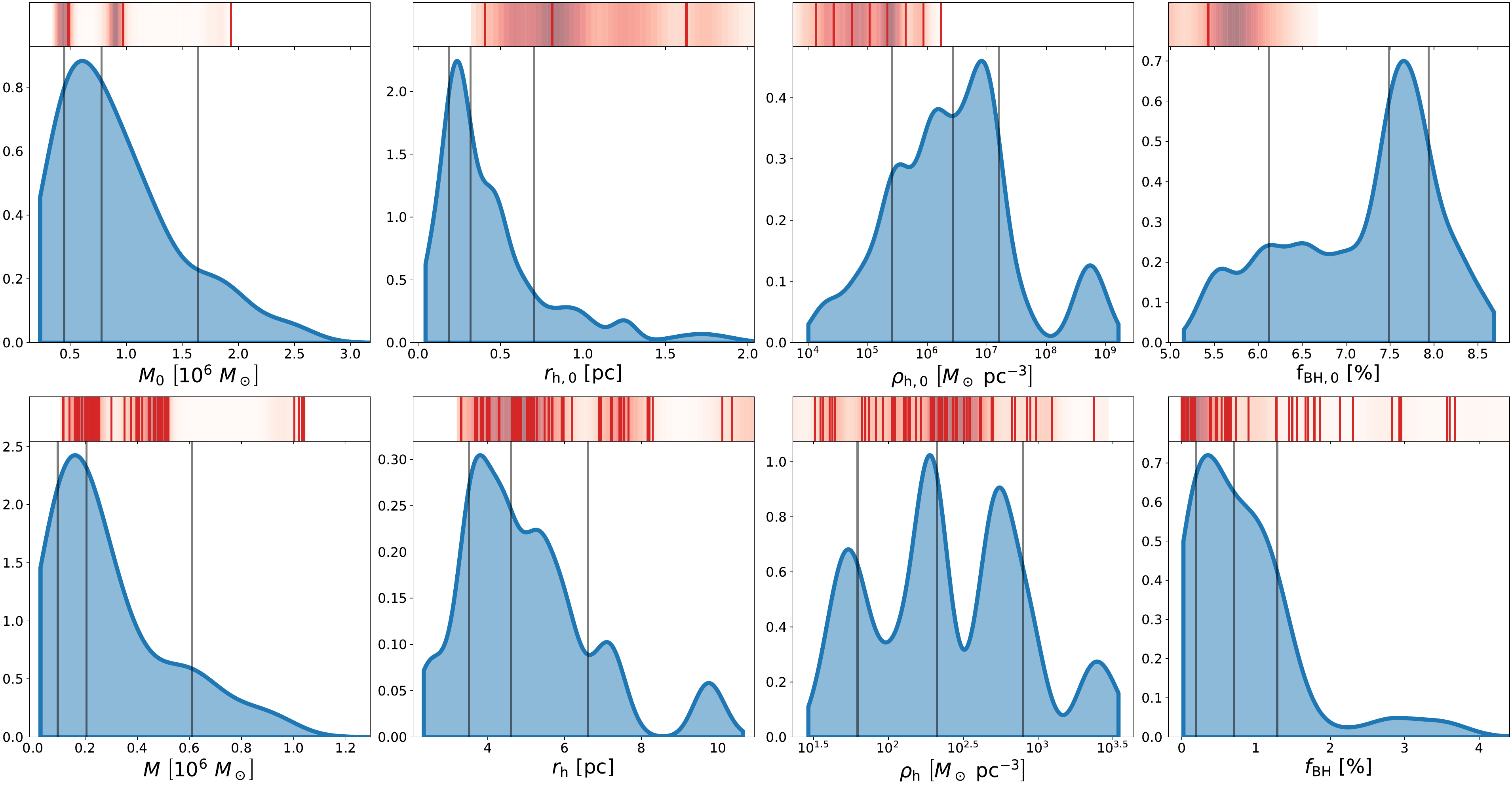}
        \caption{
            Gaussian kernel density estimates showing the overall distributions
            of the initial (top row) and final (bottom row) total mass,
            half-mass radius, half-mass density and BH mass fractions across
            the fits to all clusters in our sample.
            Median and \(1\sigma\) intervals are shown by the black lines.
            Above each panel the corresponding values for all of the CMC models
            included in our validation sample (\Cref{sec:validation}) are shown
            as red vertical lines, while the density distributions of these
            quantities from our fits to the mock CMC models are shown as the
            red shaded regions.
            The mass and present-day \fbh axes are truncated such that \omegacen
            is not visible to the far right
            (\(\Mi=6.81\pm0.04\times10^6\,\Msun\),
            \(M=5.62\pm0.02\times10^6\,\Msun\),
            \(\fbh=8.31\pm0.06\,\%\)), and not all CMC values are visible,
            in order to better highlight the distributions of the rest of the
            sample.
        }
        \label{fig:real_param_dists}
    \end{figure*}

    In \Cref{fig:real_param_dists}, we show Gaussian kernel density estimates
    of the overall distributions of the inferred initial and present-day
    mass, radius, density and BH mass fraction, across the fits to all clusters
    in our sample.
    Above each panel, we show the values of the corresponding quantities
    from the CMC validation sample (\Cref{sec:validation}), as well as the
    distributions of our fits to these mock CMC clusters, allowing comparison
    with both the CMC grid and the expected fitting
    performance characterised in \Cref{sec:validation}.

    In particular, it is interesting to note that our inferred
    present-day cluster masses and radii (and thus densities) for the
    real clusters generally fall within the range encompassed by the CMC model
    sample. The CMC grid has been shown to roughly cover the volume of
    observed present-day cluster conditions \citep{Kremer2020},
    and indeed our inferred results for individual clusters agree closely
    with other literature studies, for example with the total masses
    and half-mass radii from \citet{Baumgardt2023}.

    When we compare the initial conditions, however, certain quantities shown
    in \Cref{fig:real_param_dists}, most notably the densities and BH mass
    fractions, diverge from the CMC models.
    The initial BH mass fractions fall towards the higher end of the CMC
    grid, and frequently exceed even the maximum values seen in the grid, by
    around 1 to 2 per cent.
    As we use the same BH kick prescriptions and IFMRs here, this difference is
    attributable both to the fact that our MW sample consists of relatively
    metal-poor GCs (there are no clusters in our
    sample nearly as metal-rich as the \(Z=0.02\) CMC models),
    and to our inferred IMF (discussed in more
    detail in \Cref{sub:initial_mass_function}), both of which lead to an
    increased initial BH mass fraction.
    On the other hand, the majority of the inferred present-day BH mass
    fractions are small, typically less than 1.5 per cent
    (discussed in more detail in \Cref{sub:black_hole_populations}).

    Most interestingly, the initial radii are typically inferred to be
    quite small (\(\lesssim \SI{0.7}{\pc}\)), and in turn the initial cluster
    densities can often be notably higher than those found in even the densest
    of CMC models.
    This is discussed in more detail in \Cref{sub:initial_conditions}.
    Importantly though, as was mentioned in \Cref{sec:validation},
    while there is a spread in how well we recover the initial
    radii (and thus density, of about 1 dex) in our mock validation fits, at
    the lowest radii tested (\(r_{\mathrm{v},0}=\SI{0.5}{\pc}\)), we tend to
    slightly \textit{overestimate} the initial radii (and thus underestimate
    the density). Therefore, it is unlikely that these very
    small initial radii and high densities are the result of a corresponding
    bias in our fitting, but instead are genuinely required to fit the
    the available data.
    It must also be noted, however, that as the \clusterBH
    models are calibrated against the CMC grid, the behaviour of our models
    beyond the density regimes covered by the grid cannot be tested
    directly, and we must trust that the physical prescriptions used
    extrapolate correctly to higher densities.
    Nonetheless, even if we are less confident in the models in this regime,
    it is also true that if the observations were consistent with initial
    conditions more closely resembling those of the CMC grid, we \textit{would}
    expect to recover those, as demonstrated in \Cref{sec:validation}.
    Therefore, the fact that the data does not prefer models in this regime is
    itself indicative of higher required initial densities.

\subsection{Black Hole Populations}\label{sub:black_hole_populations}

    In this section we explore the populations of BHs present in our sample
    of clusters today, as inferred from our
    best-fitting models, and compare our results with previous studies.
    The observational data we use provides direct constraints on both
    the phase-space distribution of visible stars and the total cluster mass,
    and thus provides an indirect way of constraining the dark mass, including
    BHs, present today.

\begin{figure*}
        \centering
        \includegraphics[width=\linewidth]{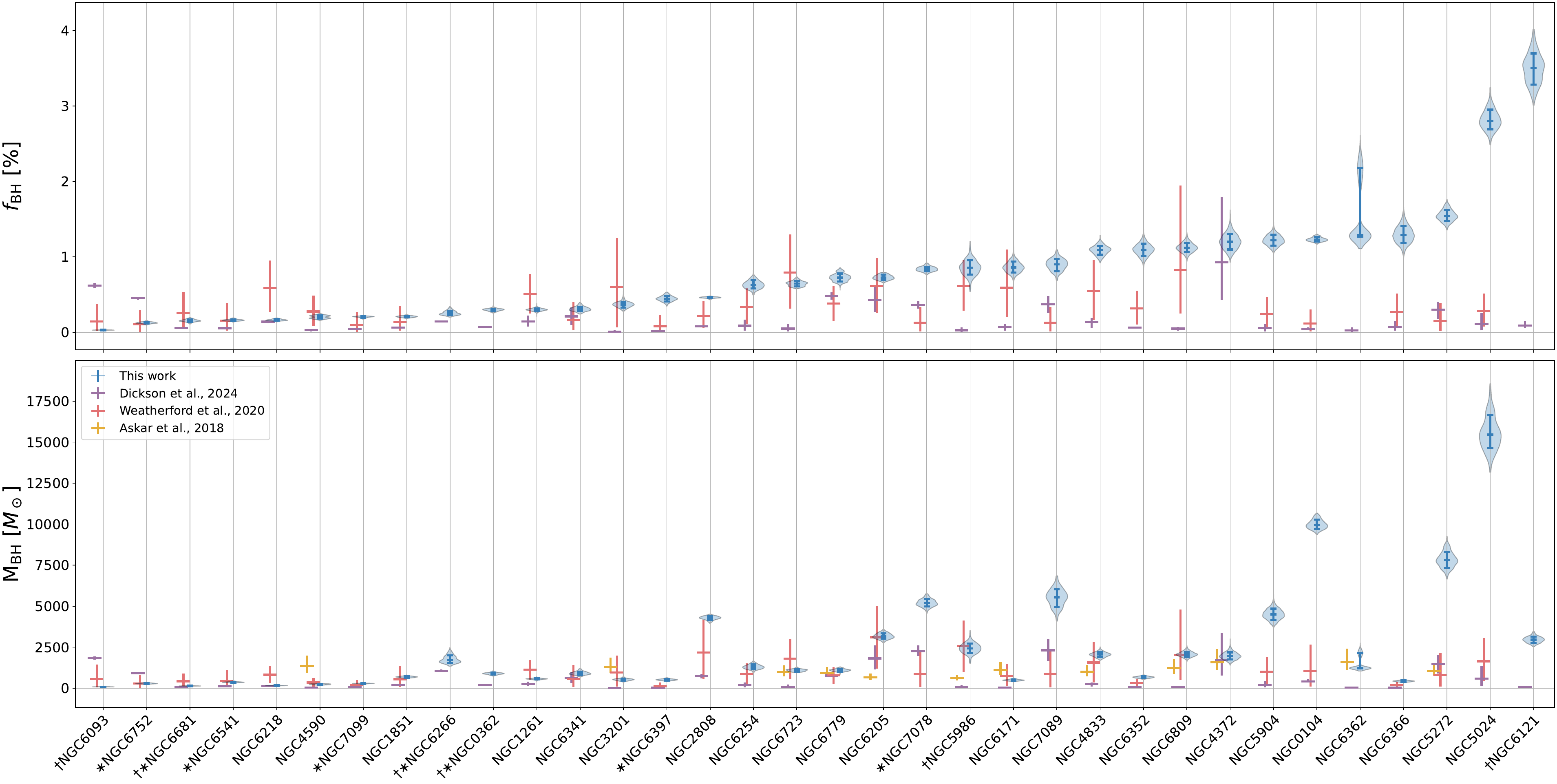}
        \caption{
            Violin plots (in blue) of the posterior probability
            distribution of the mass fraction in BHs (\fbh; upper panel) and the
            total mass in BHs (lower panel) for all clusters in our sample,
            except for \NGC{5139} (\omegacen), which has \(\fbh\sim8.3\) per
            cent, and is excluded in order to better highlight the
            distributions of the other clusters.
            The median and \(1\sigma\) intervals are denoted by
            the horizontal blue ticks within each distribution. These errors
            represent only the statistical uncertainties on our fits, and
            thus are likely underestimated.
            Clusters are sorted by \fbh.
            All clusters classified as core-collapsed in \citet{Trager1995}
            are denoted by an asterisk. All clusters with
            \(\RGeff<\SI{1.5}{\kilo\pc}\) are denoted by a dagger.
            The median and \(1\sigma\) intervals from \MMpaperII (purple),
            \citet{Weatherford2020} (red), and \citet{Askar2018} (orange),
            are also shown for all clusters in common with our sample. Values
            from \citet{Weatherford2020} are computed using
            the median clustercentric mass segregation parameter
            \(\Delta_{r50}\) (Table 1 of \citealt{Weatherford2020}),
            and conversions between total masses in BHs and mass
            fractions are computed using our total cluster mass estimates.
        }
        \label{fig:BH_violins}
    \end{figure*}

    \Cref{fig:BH_violins} shows the inferred posterior probability distributions
    of the mass fraction (\fbh) and total mass (\Mbh) in BHs retained to
    the present-day in our best-fitting models of all clusters in our sample.
    Most clusters are consistent with retaining relatively few BHs to their
    current ages, with BH mass fractions below 1 per cent in the majority
    of clusters.
    A minority of clusters have slightly higher inferred
    BH mass fractions, between 1 and 2 per cent.
    Finally, a small number of clusters have notably larger BH populations,
    greater than 2 per cent and extending up to nearly 4 per cent at the maximum
    (excluding \omegacen), though it should be noted that \NGC{6121}
    (\(\fbh\sim3.5\,\%\)) is one of the clusters with
    \(\RGeff<\SI{1.5}{\kilo\pc}\) and should be interpreted with caution.

    Also shown in \Cref{fig:BH_violins} is the comparison between our inferred
    BH masses and mass fractions and those from \MMpaperII,
    \citet{Weatherford2020} and \citet{Askar2018}.
    While, generally speaking, we reach broadly similar conclusions as in
    \MMpaperII, namely that most clusters harbour relatively small populations
    of BHs today, our inferred BH mass fractions are consistently higher
    than in \MMpaperII. We find a number of clusters with \(\fbh\gtrsim 1\)
    per cent, with the bulk of our results (\(\sim 85\) per cent) lying below
    \(\fbh=1.5\) per cent, whereas \MMpaperII instead found only one cluster
    with \fbh above 1 per cent (\omegacen), with most below \(\fbh=0.4\) per
    cent.
    On an individual level, very few of the clusters are mutually consistent
    between \MMpaperII and here, within their uncertainties,
    except where both infer very small \fbh.
    We note again, however, that only the statistical uncertainties are
    reported in both studies, and the true uncertainties are likely larger.
    Based on the results of \Cref{sec:validation}, the present-day
    \fbh values may be expected to be accurate to within 1 per cent at
    \(\fbh\sim 2.5\) per cent, at worst. While this is a rough estimate,
    it illustrates how underestimated the reported uncertainties likely are.

    \citet{Weatherford2020} predicted the amount of BHs in these clusters today
    by comparing the level of visible mass segregation against the
    anticorrelation with \fbh seen in CMC models by \citet{Weatherford2018},
    while \citet{Askar2018} used correlations found by
    \citet{ArcaSedda2018} between the density of inner BH-subsystems and the
    central surface brightness of the clusters in the MOCCA survey database.
    Given to the larger uncertainties in these studies, many of our clusters
    agree, within \(2\sigma\), with the present-day masses in BHs inferred by
    these studies, though a number of discrepancies remain
    between specific clusters, and in particular, as with the comparison to
    \MMpaperII, there are a number of clusters for which we now infer higher
    mass fractions in BHs.


    It is worth highlighting the differences in approach between \MMpaperII
    and this work in recovering the present-day BH populations.
    While the inference of the remnant populations in both are dictated by the
    total and visible mass constraints set by the observational data, and
    the same family of equilibrium models are used to fit the present-day
    structure, in \MMpaperII the final amount of BHs retained was almost
    entirely set by a freely varying BH retention fraction parameter
    (\texttt{BHret}), which controlled the total amount
    of dynamically ejected BHs, independent of all other parameters.
    In this work, however, there is no such BH retention
    parameter; instead, the initial mass fraction in BHs is set by the IMF,
    the BH IFMR and the BH natal kicks, while the subsequent rate of dynamical
    ejections over the cluster lifetime is modelled by \clusterBH, where it is
    coupled to the energy demands of the cluster as a whole.
    Thus, the final amounts of retained BHs we infer here are
    self-consistently constrained not only by the observational data, but by the
    required evolution of the rest of the cluster properties as well.

    Overall, these extra physically-motivated constraints should provide more
    confident inferences of the possible present-day BH populations.
    However, this reduction in freedom comes at the cost of additional
    modelling assumptions, some of which are governed by uncertain physical
    prescriptions.
    We have demonstrated that our models typically recover the
    BH populations of CMC simulations well (\Cref{sec:validation});
    however, since the CMC models share the same underlying BH assumptions
    (e.g. IFMR and natal kick prescriptions),
    this validation cannot be used to say whether those assumptions are valid
    for real clusters.
    One further avenue for addressing this would be to explore a wider
    range of the assumptions underlying BH formation, while still
    self-consistently modelling the evolution of the BH population.
    This is discussed in \Cref{sub:implications_for_bh_physics} and will be
    explored in more detail in a forthcoming paper.
    The fits to the observational data are similarly good in both this work and
    \MMpaperII, even though the best-fitting models may diverge from one
    another slightly in the central velocity dispersions
    (particularly when we infer larger BHs amounts), where signatures of BH
    populations should appear.
    However, the leverage provided by the BH signatures in the central
    kinematics is highly dependent on the quality and quantity
    of the central kinematic data available, and in most clusters the current
    data is insufficient to confidently distinguish the presence of BHs to a
    precision of \(\fbh\lesssim 0.5\) per cent at \(\fbh\sim1\) per cent,
    as would be required to truly
    discriminate between our results and those of \MMpaperII in most clusters,
    given we know that the overall uncertainties in both are underestimated.


    One cluster which may possess good enough kinematic data to better
    constrain its BH population is \omegacen \citep[e.g.][]{Zocchi2019},
    especially with the addition of new, deep kinematic data
    from \citet{Haberle2025}. However, the presence of several fast-moving
    stars in this new dataset has also been used to argue for the presence of a
    \(\gtrsim10^4\,\Msun\) IMBH in the cluster core
    \citep{Haberle2024}. As discussed in \MMpaperII, we cannot currently
    directly include the potential of an IMBH in our \limepy models to test
    this, and it is unclear how the presence of an IMBH would affect our
    evolutionary modelling, but our inferred mass in BHs
    (\(M_{\mathrm{BH}}=2.51\substack{+0.003 \\ -0.002}\times 10^5\,\Msun\))
    is significantly higher than the IMBH mass range reported by
    \citet{Haberle2024}, which still suggests the presence of a
    significant population of stellar-mass BHs in the cluster.
    We also note that, given the possible origin of \omegacen as the
    accreted nuclear star cluster of a disrupted dwarf galaxy
    \citep[e.g.][]{Bekki2003,Meza2005}, the tidal history of this cluster may
    be more complex than can be accounted for in our models.

\subsection{Initial Mass Function}\label{sub:initial_mass_function}

    In this section we explore the shape of the stellar IMF inferred from
    our fits.
    We place direct constraints on the IMF below \SI{1}{\Msun} by
    allowing the power-law slopes of \(\alpha_1\) and \(\alpha_2\) to vary
    freely.

\begin{figure}
        \centering
        \includegraphics[width=\linewidth]{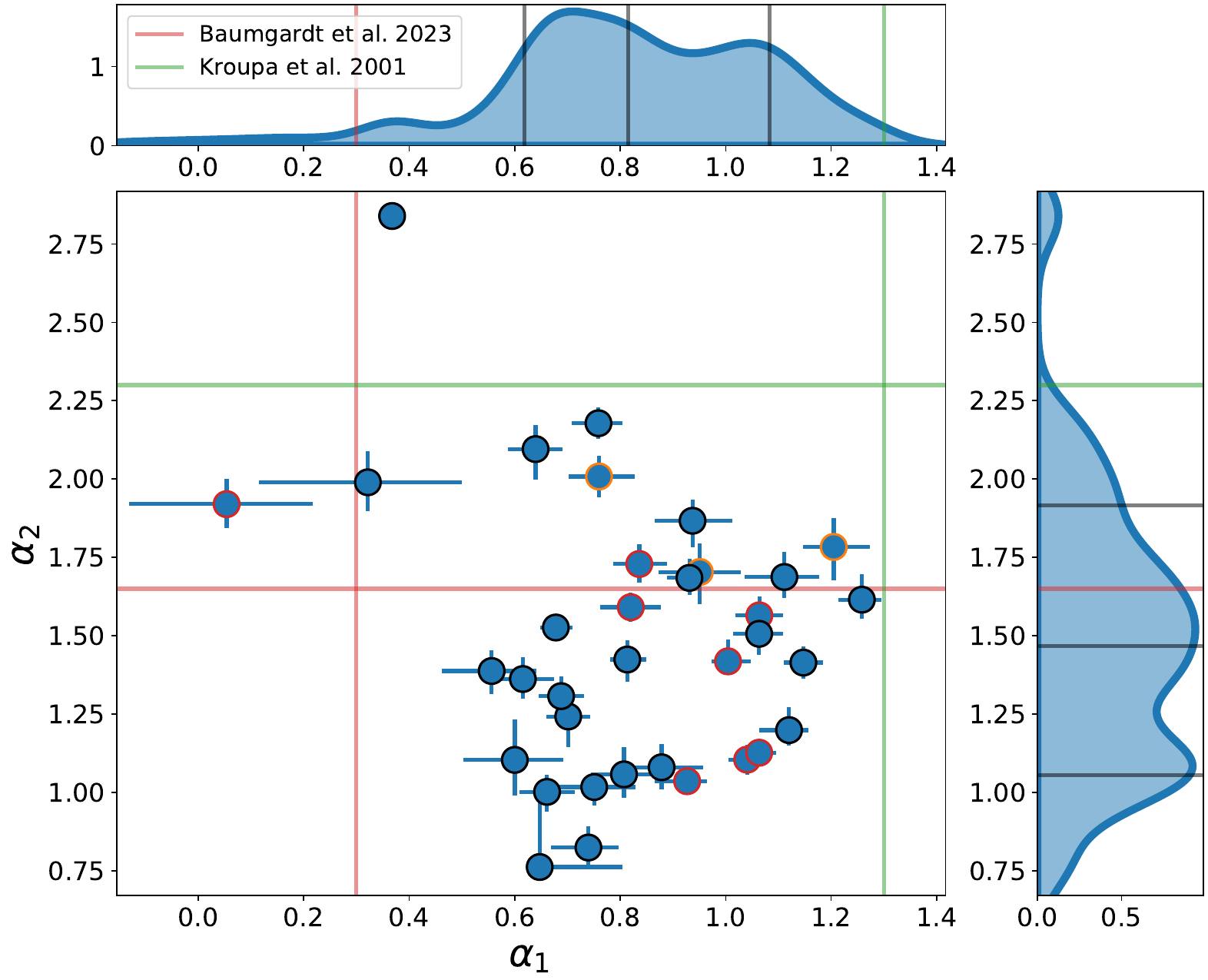}
        \caption{
            The median values and \(1\sigma\) uncertainties of the inferred
            stellar IMF power-law slopes \(\alpha_1\) and \(\alpha_2\) for
            all clusters in our sample.
            Gaussian kernel density estimates showing the overall distributions
            of each parameter are shown above and to the right.
            The median values of the corresponding IMF slopes from
            \citet{Kroupa2001} (green) and \citet{Baumgardt2023} (red)
            are shown by the horizontal and vertical lines on each panel.
            All core-collapsed clusters are outlined in red.
            All clusters with \(\RGeff<\SI{1.5}{\kilo\pc}\) are outlined in
            orange.
        }
        \label{fig:IMF_slopes}
    \end{figure}

    In \Cref{fig:IMF_slopes} we show the inferred values of the IMF slopes
    below \SI{1}{\Msun}.
    This figure clearly shows that our inferred IMF does not match
    that of \citet{Kroupa2001}, a commonly adopted IMF prescription which is
    used in most cluster modelling, including the CMC grid itself.
    Our inferred IMF is notably more bottom-light, i.e. deficient
    in low-mass stars, relative to the \citet{Kroupa2001} IMF. This is in line
    with the findings of \citet{Baumgardt2023}, though as also shown in
    \Cref{fig:IMF_slopes} our inferred slopes differ from theirs as well.
    Over our full MW cluster sample we find median and \(1\sigma\) values of
    \(\alpha_1= 0.82\substack{+0.27 \\ -0.20}\) and
    \(\alpha_2= 1.47\substack{+0.45 \\ -0.41}\), in comparison with the values
    of \(\alpha_1=0.3\) and \(\alpha_2=1.65\) for \citet{Baumgardt2023} and
    \(\alpha_1=1.3\) and \(\alpha_2=2.3\) for \citet{Kroupa2001}.
    We note, however, that \citet{Baumgardt2023} use a slightly
    different break mass in this regime of \SI{0.4}{\Msun} compared to our
    \SI{0.5}{\Msun}, and an IMF upper mass limit of \SI{100}{\Msun} compared
    to our limit of \SI{150}{\Msun}.
    These slopes are also similar to those found for the least
    dynamically evolved clusters in \MMpaperI (see their Figure 8).
    The slopes inferred there effectively represent the present-day
    mass function rather than the true IMF, as no prescriptions were employed
    to account for the tidal losses of stars, which preferentially
    affects the low-mass stars and changes the shape of the mass function over
    time; however, in the least dynamically evolved
    clusters, the present-day MF should resemble the IMF most closely.

    In Appendix~\ref{sec:cmc_massfunc_comp} we demonstrate how these inferred
    IMF slopes are mainly constrained by the observed stellar mass function
    data.

    This bottom-light IMF has several important implications, in comparison
    to canonically assumed IMF prescriptions such as \citet{Kroupa2001}.
    This IMF yields a notably higher initial mean stellar mass
    (\(\sim \SI{0.97}{\Msun}\), compared to \SI{0.58}{\Msun} under
    \citealt{Kroupa2001}).
    It also results in a nearly 25 per cent increase in the number of BHs
    formed, per unit cluster mass.
    This is largely the reason why, as noted in \Cref{sec:real_fitting},
    our best-fit models typically have higher initial BH mass fractions
    (peaking near \(\fbhi\simeq 7.7\%\)) than in, e.g., the CMC grid.

\subsection{Initial Conditions}\label{sub:initial_conditions}

    One of the most important results enabled by our new models is the
    inference of the clusters’ initial conditions: total mass, half-mass
    radius and density.
    As described in \Cref{sec:methods} and tested in \Cref{sec:validation},
    we directly infer these quantities as free parameters in our fitting,
    constrained by the observational constraints on their present-day
    counterparts, and the evolutionary history implied by \clusterBH.

    \begin{figure}
        \centering
        \includegraphics[width=\linewidth]{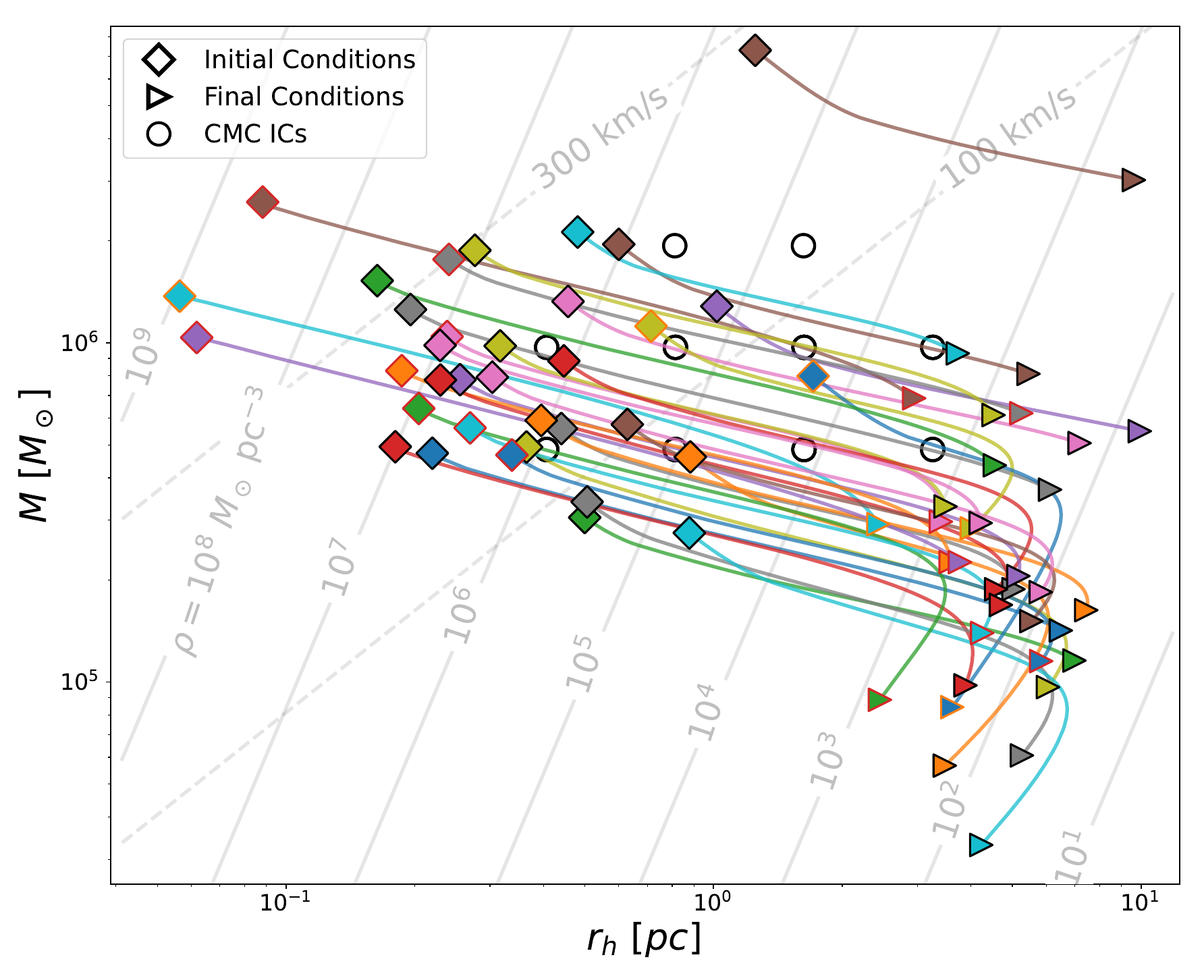}
        \caption{
            Tracks of the evolution of each cluster in our sample, from
            initial total cluster mass and half-mass radius (square markers)
            over their lifetimes to the present conditions (triangle markers).
            Only the median inferred values of each quantity are shown.
            All colours are assigned randomly, to differentiate the tracks.
            Lines of constant half-mass density and central escape velocity
            are shown in grey.
            The initial values of the sample of CMC models used in
            \Cref{sec:validation} are also shown as circles.
            All core-collapsed clusters are outlined in red.
            All clusters with \(\RGeff<\SI{1.5}{\kilo\pc}\) are outlined in
            orange.
        }
        \label{fig:ev_tracks}
    \end{figure}

    The distributions of the inferred initial conditions across our fits are
    shown in the top row of \Cref{fig:real_param_dists},
    while \Cref{fig:ev_tracks} shows the median initial cluster mass and
    half-mass radius for each cluster, as well as their evolutionary tracks
    over the age of the cluster to their present-day values.
    The individual quantities for each cluster are also listed in
    \Cref{table:best_fitting_params,table:extra_params}.

    The majority of our clusters have initial total masses within
    a range of \(0.4 - 2 \times 10^6\, \unit{\Msun}\),
    while the initial radii are typically less than \SI{0.7}{\pc},
    with the overall distribution peaking around \SI{0.25}{\pc}.
    In many clusters, these compact initial sizes result in very high initial
    densities.
    The overall distribution of \rhohi has a median value of about
    \(10^{6.4}\, \unit{\Msun \ \pc^{-3}}\) with a \(1\sigma\) width of slightly
    less than 1 dex.
    A notable peak can also be seen at much higher densities
    (\(\sim10^{9}\, \unit{\Msun \ \pc^{-3}}\)) in \Cref{fig:real_param_dists},
    however this is made up of clusters with \(\RGeff<\SI{1.5}{\kilo\pc}\)
    (\NGC{6093} and \NGC{6266}), which should be regarded with caution, as
    discussed in \Cref{sub:outliers}.


    Also overlaid in \Cref{fig:ev_tracks} are the initial values of the CMC
    models used in \Cref{sec:validation}. A significant fraction of our
    clusters have inferred initial densities exceeding even the densest
    CMC models.
    As discussed in \citet{Kremer2020}, this may partly be simply due to the
    limits of their simulation grid,
    due to the high computational costs and stellar collision rates of
    the densest model.


    It is important to note that the distributions shown here
    of, e.g., initial density reflect only the inferred conditions of the
    individual clusters within our sample.
    The initial conditions of the individual clusters are not drawn from
    shared parent distributions (e.g. a common cluster mass function) but
    from independent, uninformative priors, and do not account for, e.g.
    the populations of initially formed clusters which have dissolved by the
    present day.
    Our sample of clusters is also likely biased towards more nearby, massive
    and denser GCs, and thus not necessarily representative of the entire
    surviving MW GC population.
    Therefore, these distributions should not be taken as overall cluster
    population distributions, but only to demonstrate where the initial
    conditions of a large sample of the most well-studied MW clusters may lie.
    The fast models presented here are, however, well
    suited to this problem, and in a future work we will explore the
    cluster population distributions directly using
    hierarchical Bayesian modelling approaches.



\section{Discussion}\label{sec:discussion}

\subsection{Comparison with High Redshift Clusters}
\label{sub:high_redshift}

\begin{figure}
        \centering
        \includegraphics[width=\linewidth]{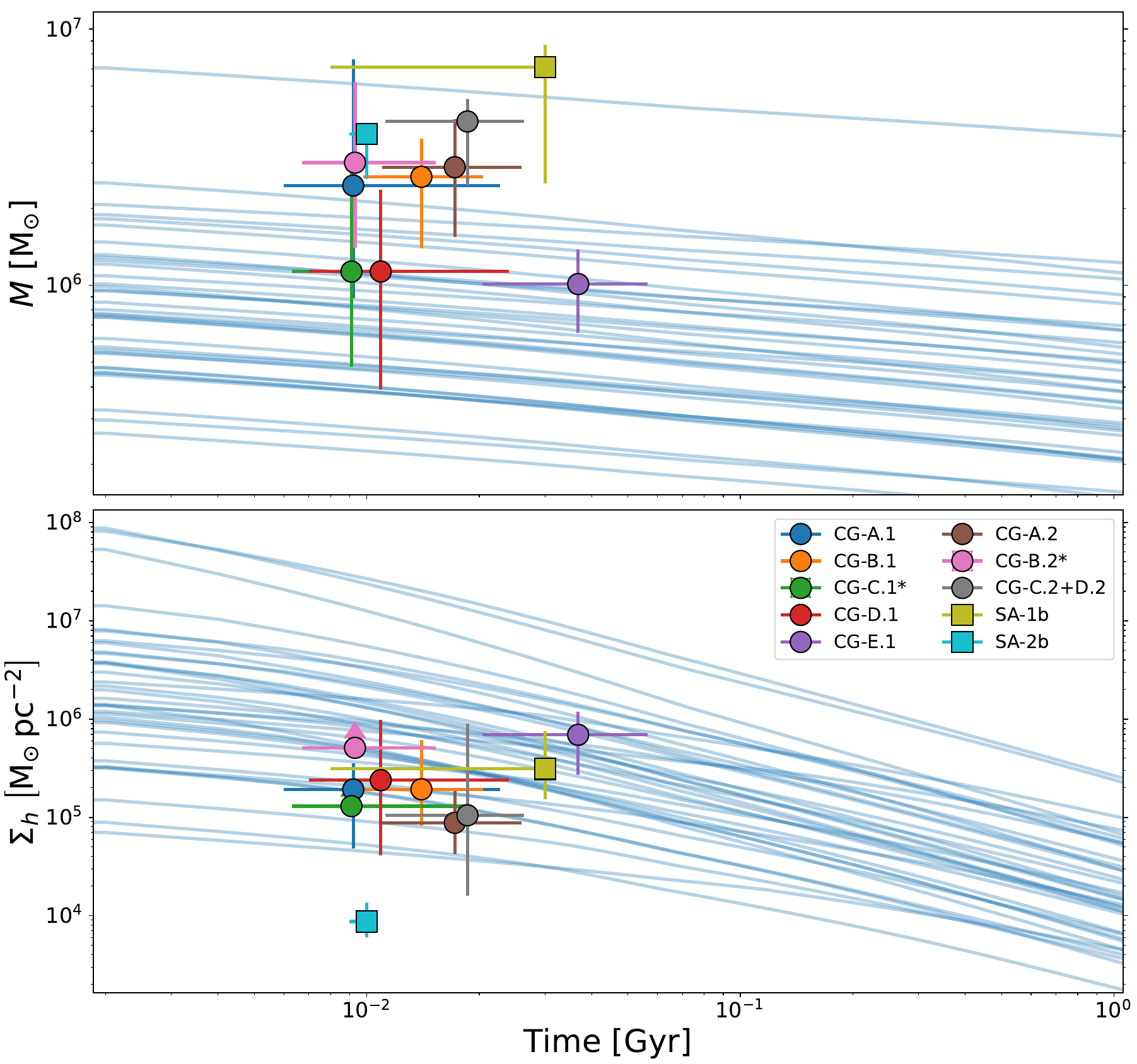}
        \caption{
            Evolution of the inferred total mass (top panel) and
            average effective surface density (bottom panel) over
            time for all clusters in our sample (blue lines). Only the median
            inferred values are shown. The surface density is computed assuming
            that light follows mass, and by scaling the half-mass
            radius by 0.75, to approximately account for projection effects.
            The circle and square markers represent the age, mass and
            surface density measurements of the ``Cosmic Gems'' (CG) and
            ``Sunrise Arc'' (SA) proto-GCs as presented by
            \citet{Adamo2024} (see their Table 1 and Extended Data Table 2)
            and \citet{Vanzella2023} (see their Table 3, excluding those
            classified as ``nebular knots''), respectively.
            Clusters denoted with an asterisk represent only lower limits on the
            density.
        }
        \label{fig:gems_comp}
    \end{figure}

    With the advent of JWST, it is now becoming possible to compare
    our modelling predictions with observations of high-redshift
    candidate proto-GCs in strongly lensed galaxies.
    In \Cref{fig:gems_comp}, we compare the young GC-like systems discovered
    by \citet{Adamo2024} within the ``Cosmic Gems Arc'' and \citet{Vanzella2023}
    within the ``Sunrise Arc'' with
    the inferred density evolution of our full sample of MW GCs.
    The bulk of our results are consistent with these observations.
    All but the densest outliers reach comparable densities to these
    high-redshift clusters by their estimated ages
    (\(\sim 20 - \SI{50}{\Myr}\)). This is possible, despite many of our
    inferred initial densities being higher than those observed, because within
    the first \(\sim \SI{100}{\Myr}\) the density of a cluster can decrease by
    nearly an order of magnitude as it quickly loses mass and expands.
    This comparison suggests that our inferred initial densities for MW GCs
    may be consistent with proto-GCs observed in the early Universe.
    However, there also remains substantial uncertainties in the
    physical parameters of these candidate proto-GCs, due to uncertainties in
    the stellar population and lens modelling \citep{Adamo2024}.

\subsection{Runaway Collisions and IMBH Formation}

    Here we discuss the consequences of the high initial densities we infer
    on the formation of an IMBH from a very massive star
    (\(\gg\SI{100}{\Msun}\)) formed through runaway stellar collisions
    \citep[e.g.,][]{PortegiesZwart2004,Fregeau2004,Sharma2025}.
    Apart from \omegacen \citep{Haberle2024}, there is currently no
    compelling evidence for IMBHs in any MW GCs, with current upper
    limits of \(\lesssim10^3\,\unit{\Msun}\) from both dynamics
    \citep[e.g.,][]{Baumgardt2017} and (the absence of) accretion signatures
    \citep[e.g.,][]{Termou2018}.

    As a method of estimating the expected IMBH masses which could arise
    given our densities, we can turn to the results of Monte Carlo and \Nbody
    simulations by \citet{Vergara2026}.
    These authors demonstrated that above a critical cluster
    mass (\(M_{\mathrm{crit}}\)), runaway collisions can lead to IMBHs with
    masses \(\gtrsim10^4\,\unit{\Msun}\).
    This \(M_{\mathrm{crit}}\) depends on the initial radius as
    \(M_{\mathrm{crit}}\propto \rh^{7/3}\), and if we rewrite this in terms of
    \rhoh using their constants of proportionality, we find that the critical
    \textit{density} for formation of IMBHs with masses
    \(\gtrsim10^4\,\unit{\Msun}\) is
    \begin{equation}
        \rho_{\mathrm{h,crit}} \gtrsim 10^8\,\unit{\Msun \, \pc^{-3}}\,
            \left(\frac{M}{10^6\,\unit{\Msun}}\right)^{-2/7}.
    \end{equation}
    All of our inferred initial densities (except for the three highest-density
    outliers, which as discussed above should all be taken with some caution),
    fall below this mass-dependent \(\rho_{\mathrm{h,crit}}\).
    \citet{Rantala2026} similarly explored the formation of IMBHs via runaway
    collisions within a grid of \Nbody models spanning a range of densities.
    Based on these simulations, the authors presented simple formulae for
    the expected IMBH masses as a function of cluster surface density and
    metallicity (see their Equations 9 and 10).
    Applying these functions to our inferred surface densities results in
    IMBH masses in the range of around 1000 to \SI{3500}{\Msun}.
    Comparison with these studies thus suggests that, even given our notably
    higher inferred initial densities, we might not expect to see very massive
    IMBHs (\(\gtrsim10^4\,\Msun\)) form in GCs through this mechanism.

    \citet{Bocchi2026} presented another method for estimating the potential
    IMBH masses in dense clusters by employing Fokker-Planck models, and
    applied this to high-redshift clusters.
    For the COSMIC Gems and Sunrise proto-GCs (\Cref{sub:high_redshift}), they
    estimated IMBH masses of around
    \((1-3)\times10^3\,\unit{\Msun}\). Given that many of our clusters match
    the densities of these proto-GCs, this may suggest that similar IMBH masses
    could form within them as well.

    There are, however, a number of caveats related to this analysis.
    First, stellar collisions may be far more efficient if the IMF were to
    extend up to much larger masses (e.g. \(\sim10^3\,\unit{\Msun}\)), as
    was recently proposed by \citet{Gieles2025} to explain the multiple stellar
    population of GCs.
    On the other hand, collisions between stars and very/extremely massive
    stars are not well studied, and significant uncertainties remain.
    For example, these collisions may actually lead to a net stripping of mass
    \citep{RamirezGaleano2025,RomanGarza2026}.
    Whether these collisions add or remove mass depends sensitively on
    the structure of very/extremely massive stars, which is itself also
    very uncertain.
    Furthermore, as noted in \Cref{sub:initial_conditions}, our own
    evolutionary models come with the caveats that we do not account for these
    collisions or IMBH formation.

    In summary, given our high inferred initial densities, we may expect
    stellar collisions to occur and potentially lead to the formation of IMBHs
    with masses \(\lesssim10^3\,\unit{\Msun}\). However, it is not possible for
    us to place strong constraints on this process.

\subsection{IMBH Formation Through Hierarchical BH Mergers}

    An alternative pathway to IMBH formation in
    GCs is through repeated, or ``hierarchical'', BH mergers
    \citep[e.g.,][]{Antonini2016, Rodriguez2018, Mapelli2021}
    Here we consider the implications of our inferred initial conditions for
    IMBH formation via hierarchical BH mergers.

    Whether this occurs depends critically on the
    initial escape velocity of the GC (\vesci), which must be high
    enough to retain the first few BBH merger products after the
    gravitational wave (GW) recoil kicks they receive during each merger.
    These kicks are particularly large for near equal-mass mergers
    (hundreds of \unit{\km\per\second}). Subsequent mergers have smaller mass
    ratios and receive smaller GW kicks, allowing the most massive BH
    to continue to grow.
    \citet{Antonini2019} showed that an IMBH could grow in this way in
    GCs with \(\vesci\gtrsim\SI{300}{\km\per\second}\).
    The final IMBH mass then scales nearly linearly with GC mass
    as \(M_{\mathrm{IMBH}}\simeq 2\times10^{-3}\,M\), with a weak dependence on
    initial density (see their equation 21).
    Similar conclusions were reached by \citet{Torniamenti2026}.

    \Cref{fig:ev_tracks} shows that three of our clusters (\NGC{6093},
    \NGC{6266} and \NGC{6752}) have \(\vesci\gg\SI{300}{\km\per\second}\),
    however, as noted in \Cref{sub:outliers}, each of these three should be
    treated with caution.
    Five other clusters (\NGC{5139}, \NGC{5904}, \NGC{6205},
    \NGC{7078}\(^\ast\), \NGC{7089}\(^\ast\)) have initial escape velocities
    approaching this limit (\(\vesci>\SI{250}{\km\per\second}\)).
    Of these, only \NGC{5139} (\omegacen) has compelling evidence for an IMBH,
    and the expected IMBH mass in the hierarchical merger scenario
    (\(\sim1.6\times10^4\,\Msun\)) is consistent with the inferred IMBH mass of
    \citet{Haberle2024}.
    For the other GCs, no conclusive evidence of IMBHs has been presented,
    though expected IMBH masses of a few times \(10^3\,\Msun\) in
    lower-mass GCs cannot always be ruled
    out \citep[e.g.,][]{McNamara2012,denBrok2014}.

    In summary, our inferred initial conditions suggest that hierarchical
    BH mergers could present a viable avenue for IMBH growth in a small
    number of MW GCs, however for the majority of clusters
    (where \(\vesci<\SI{300}{\km\per\second}\)), this channel is unlikely.

\subsection{Maximum Allowed Surface Densities}

    It has been suggested, based on the maximum density observed in
    dense stellar systems today across many orders of magnitude
    \citep{Hopkins2010} and on theoretically expected equilibrium states
    between feedback and gravity \citep{Crocker2018,Grudic2019}, that there
    is an upper limit expected on the initial stellar surface
    densities of newly formed star clusters.
    While the proposed maximum values vary slightly, a rough upper limit
    slightly above of \(\sim10^5\,\unit{\Msun \ \pc^{-2}}\) has been proposed.
    For nearly every cluster in our sample, our inferred initial surface
    densities clearly exceed these proposed limits.
    While our modelling does have the caveat
    that we simplify the very beginnings of our cluster evolution
    by assuming all of the surrounding gas has been cleared (at time 0),
    it is nonetheless true that much higher initial stellar densities are
    clearly able to recreate the GCs we observe today, and are in fact
    preferred by the data.
    Similarly, as discussed in \Cref{sub:high_redshift}, more recent
    observations of high-redshift proto-GCs may also lie above this
    limit, even after \(\sim \SI{50}{\Myr}\) of evolution
    \citep{Adamo2024,Vanzella2023}.

\subsection{Implications for Gravitational Wave Sources}

    The density of star clusters is an important parameter in setting the
    rate of dynamically-formed BBH mergers, with denser
    clusters generating more BBH mergers \citep[e.g.,][]{Rodriguez2016}.
    Here we consider whether our inferred initial densities are consistent with
    the BBH merger rate determined by LIGO-Virgo-KAGRA (LVK).
    The population models of \citet{Antonini2020b} show that the dynamical
    BBH merger rate at low redshift increases with initial GC density up to
    densities of \(\sim10^5\,\unit{\Msun \, \pc^{-3}}\), beyond which the
    merger rate plateaus and even decreases slightly for
    densities \(\gtrsim10^7\,\unit{\Msun \, \pc^{-3}}\) as BBH mergers
    in denser GCs tend to occur at higher redshifts.
    The maximum dynamical merger rate in their models is well below the total
    merger rate inferred by LVK, and therefore our inferred densities
    do not imply an overproduction of dynamical BBH mergers.

    Denser clusters also give rise to a steeper redshift dependence for the
    merger rate \citep[see Figure 4 of][]{Antonini2020b}, with densities of
    \(\sim10^6-10^7\,\unit{\Msun \, \pc^{-3}}\) (in line with our results)
    leading to a redshift dependence consistent with recent LVK findings
    \citep[GWTC-5.0;][]{LIGO2026}.

    Finally, initial densities of \(\gtrsim10^5\,\unit{\Msun \, \pc^{-3}}\) are
    needed to explain the most massive BHs in the pair-instability gap with
    hierarchical mergers in GCs \citep{Antonini2023}.

    In summary, population models of GCs seem to favour initial densities
    similar to those we infer here, in order to explain some features of the
    observed GW data.

\subsection{Implications for BH Physics}
\label{sub:implications_for_bh_physics}

    The high initial densities we infer could also instead reflect underlying
    issues in some of the prescriptions used within our models, and
    which are also commonly assumed in many other dynamical models.
    The constraints on the present-day conditions from the observations,
    notably the central kinematics and stellar mass functions, require
    both an IMF that is deficient in low-mass stars, and thus
    enriched in BHs progenitors
    (as discussed in \Cref{sub:initial_mass_function}; see also \MMpaperI and
    \citealt{Baumgardt2023}), and in most clusters relatively small populations
    of BHs retained to the present day (as discussed in
    \Cref{sub:black_hole_populations}).
    In order to achieve this, while also reproducing the total masses and radii
    observed today, a very high rate of BH ejections from the cluster is
    required.
    Within our current framework, the only viable method for
    accomplishing this is to increase the rate of early dynamical ejections by
    substantially increasing the initial densities and shortening the
    relaxation times.

    However, this is not the only way by which the large required depletion of
    the initial BH population could be achieved.
    The BH natal kicks also provide a mechanism for ejecting the BHs immediately
    after their formation.
    Stronger kicks could allow a cluster to deplete the
    populations of BHs and yield the required present-day conditions,
    without requiring a higher initial density.
    Similarly, a different BH IFMR could reduce the initial mass in BHs
    formed from our bottom-light IMF, alleviating this issue.
    The BH natal kick and IFMR prescriptions that we use here (as described in
    \Cref{sec:methods}) are both commonly used by many other cluster modelling
    works, but they remain uncertain \citep{Boccioli2024,Popov2025}.

    To demonstrate this, we conduct a simple experiment using \clusterBH.
    We start with a fiducial set of initial conditions
    (\(\Mi=10^6\ \unit{\Msun}\), \(\rhohi= 10^6\ \unit{\Msun \, \pc^{-3}}\),
    \(\FeH=-2\), \(\RGeff=\SI{5}{\kilo\pc}\) and the median inferred IMF slopes
    from \Cref{sub:initial_mass_function}) and,
    instead of the canonical kick prescription used in this paper, we reduce
    the initial BH mass by a fraction \(\fkick=50\%\), and then evolve this
    model for \SI{12}{\Gyr}.
    We then consider a grid of initial densities, spanning an order
    magnitude in either direction (\(\rhohi\in[10^5,\,10^7]\,
    \unit{\Msun \, \pc^{-3}}\)), hold all other parameters fixed, and determine
    for each density the value of \fkick that leads to an evolution which
    minimizes the distance between the present-day values of the total mass,
    radius, and BH mass fraction of the fiducial and modified models.
    This yields a relationship between the initial density and the strength of
    the natal kicks which produces a nearly identical cluster at the present
    day.

\begin{figure}
        \centering
        \includegraphics[width=\linewidth]{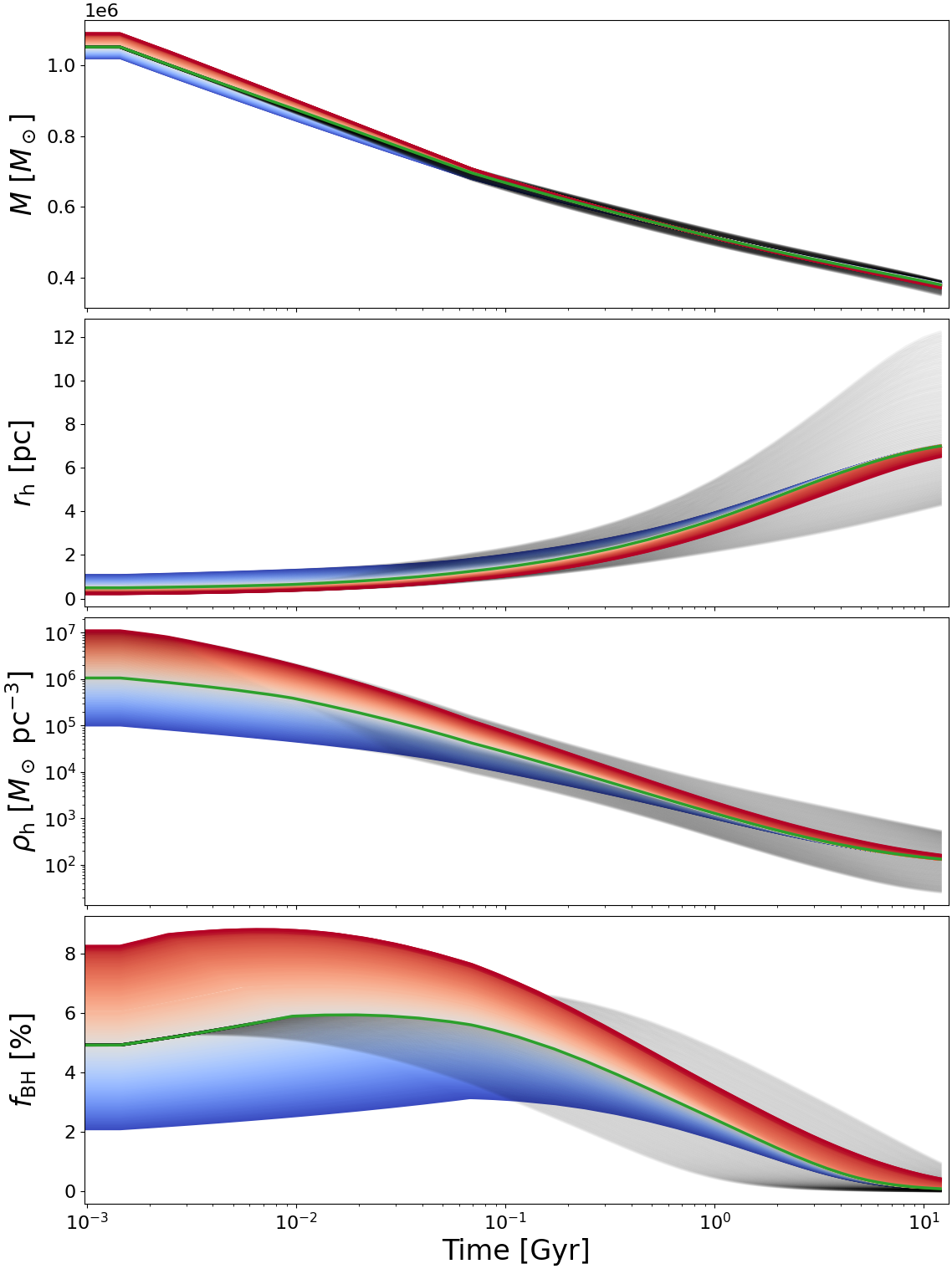}
        \caption{
            Evolution of cluster total mass, half-mass radius, half-mass density
            and BH mass fraction over time for a fiducial \clusterBH model with
            initial conditions \(\Mi=\SI{1e6}{\Msun}\),
            \(\rhohi= \SI{1e6}{\Msun \, \pc^{-3}}\), \(\FeH=-2\),
            \(\RGeff=\SI{5}{\kilo\pc}\), \(\alpha_1=-0.82\) and \(\alpha_2=-1.45\)
            (green line), as well as a set of models evolved from a grid of
            initial densities and corresponding BH natal kick strengths (\fkick)
            which result in similar present-day conditions (red to blue lines).
            The grey background lines represent models over the same grid of
            initial densities with a fixed \(\fkick=50\%\), showing how
            their evolution differs under the same BH kick prescription.
        }
        \label{fig:fkick_experiment_ev}
    \end{figure}

    The results of this experiment are shown in \Cref{fig:fkick_experiment_ev},
    where we can see that a change in kick strength of approximately 40 per
    cent, in either direction, can compensate for
    an order-of-magnitude change in initial density in the opposite direction,
    and result in very similar models at the present day.
    This demonstrates the existence of a clear degeneracy between the initial
    mass in BHs and the initial cluster density. It also shows that if the
    BH natal kicks are in reality stronger than commonly assumed,
    lower initial cluster densities may also be consistent with
    present-day observations.
    In a forthcoming paper we will harness the flexibility of our
    models to test this hypothesis in more detail by refitting our
    sample of clusters under different BH prescriptions and examining the
    impacts on the inferred initial conditions.

\section{Conclusions}\label{sec:conclusions}

    In this work, we have combined fast, evolutionary cluster models
    with multimass equilibrium models, providing a novel tool for the
    inference of the initial conditions of GCs from observations of
    their present-day structure and kinematics.
    We have validated these new coupled models and their ability to recover
    both present-day and initial cluster conditions against mock observations
    of a large grid of CMC models.
    Finally we have applied them to a sample of 35 MW clusters,
    fitting to a number of observed proper motion, line-of-sight velocity,
    number density, and stellar mass function datasets, allowing us to
    explore the present-day stellar and remnant populations, and place
    constraints on their initial conditions.
    This has yielded a number of interesting conclusions:

    \begin{enumerate}

        \item
        We infer a distribution of initial half-mass densities across our sample
        which is notably high, peaking at \(\log_{10}\left(\rhohi\right) =
        6.44\substack{+0.72 \\ -1.03}\), driven by typically very compact
        values of \rhi.

        \item
        In line with recent studies (e.g. \MMpaperI; \citealt{Baumgardt2023}),
        we recover a stellar IMF which is \textit{bottom-light} (i.e.
        deficient in low-mass stars) in comparison with canonical IMF
        prescriptions, with low-mass power-law slopes of
        \(\alpha_1= 0.82\substack{+0.27 \\ -0.20}\) and
        \(\alpha_2= 1.47\substack{+0.45 \\ -0.41}\).

        \item
        The majority of our GCs are consistent with having retained relatively
        small populations of BHs to the present day (typically less than
        \(\fbh=1.5\) per cent), despite initially higher BH mass fractions,
        arising from our inferred IMF.

        \item
        Despite the high inferred initial densities, many of our clusters are
        consistent with the recent density estimates of candidate
        proto-GCs at high redshifts, such as the COSMIC Gems clusters
        \citep{Adamo2024}.

    \end{enumerate}

    These results, especially the notably high initial densities
    we infer, have important implications for many open fields of
    study, ranging from cluster formation to BH growth to GW sources.

    These results
    may also be, however, particularly sensitive to the exact BH formation
    prescriptions assumed (e.g. their IFMRs and supernova kicks), and different
    assumptions could lead to remarkably different inferred initial conditions.
    While we adopt commonly assumed recipes here, these remain very
    uncertain and we have demonstrated that degeneracies exist
    between initial conditions and the initial BH mass fractions, which are
    in large part set by the adopted BH prescriptions.
    In a forthcoming work, we will investigate this problem in more
    detail by re-fitting our sample of clusters under a variety of different
    prescriptions for the BH natal kicks and IFMRs, and exploring whether, for
    example, stronger natal kicks allow for recovering the observed properties
    of these clusters with lower initial densities.

    In addition, the distributions of initial conditions we present here
    represent only those of the individual clusters within our sample.
    To constrain the overall, shared, population-level
    distributions of initial conditions (e.g. the cluster mass function),
    accounting for the full MW GC population, including dissolved and
    undetected clusters, would be required.
    In a future work, we will tackle this problem by applying our fast
    evolutionary models within a fully hierarchical Bayesian modelling
    framework.

\section*{Acknowledgements}

    ND is grateful for the support of the Durland Scholarship in Graduate
    Research and the MITACS Globalink Research Internship program, and for the
    hospitality of ICCUB during the completion of this work.
    VHB acknowledges the support of the Natural Sciences and Engineering
    Research Council of Canada (NSERC) through grant RGPIN-2020-05990.
    FFP acknowledges the “la Caixa” Foundation (ID100010434) for financial
    support in the form of a Doctoral INPhINIT fellowship (fellowship code
    LCF/BQ/DI23/11990067). MG acknowledges financial support
    from the grants  PID2024-155720NB-I00, CEX2024-001451-M funded by
    MCIN/AEI/10.13039/501100011033 (State Agency for Research of the
    Spanish Ministry of Science and Innovation).

    This research was enabled in part by support provided by ACENET
    (\url{www.ace-net.ca}) and the Digital Research Alliance of Canada
    (\url{https://alliancecan.ca}).

    \software{
        \texttt{astropy} \citep{Astropy2018},
        \texttt{dynesty} \citep{Speagle2020,Koposov2025},
        \texttt{h5py} \citep{Collette2022},
        \texttt{cmctoolkit} \citep{Rui2021a},
        \texttt{matplotlib} \citep{Hunter2007},
        \texttt{numpy} \citep{Harris2020},
        \texttt{scipy} \citep{Virtanen2020}
    }


    \bibliographystyle{aasjournal}
    \bibliography{biblio}


\appendix

    \setcounter{table}{0}
    \renewcommand{\thetable}{A\arabic{table}}


\section{All Fit Parameters}\label{sec:all_fit_parameters}

    \begin{table}
    \renewcommand*{\arraystretch}{1.3}
    \centering
    \scriptsize
    \movetabledown=5cm
    \begin{rotatetable}
    \begin{tabular}{lccccccccccccc}
    \toprule
    Cluster & \(\hat{\phi}_0\) & \(M_{0}\ \left[10^6\ M_\odot\right]\) & \(r_{\mathrm{h},0}\ \left[\mathrm{pc}\right]\) & \(\log\left(\hat{r}_\mathrm{a}\right)\) & \(g\) & \(s^2\ \left[\mathrm{arcmin^{-4}}\right]\) & \(F\) & \(\alpha_1\) & \(\alpha_2\) & \(d\ \left[\mathrm{kpc}\right]\) & \(\delta\) & \(\zeta\) & \(\eta\) \\
    \midrule
    \NGC{104} & \(15.3\substack{+0.4 \\ -0.4}\) & \(1.87\substack{+0.06 \\ -0.04}\) & \(0.59\substack{+0.04 \\ -0.05}\) & \(5.7\substack{+1.4 \\ -0.9}\) & \(1.55\substack{+0.03 \\ -0.04}\) & \(0.02\substack{+0.03 \\ -0.01}\) & \(4.6\substack{+0.1 \\ -0.1}\) & \(0.64\substack{+0.05 \\ -0.05}\) & \(2.10\substack{+0.08 \\ -0.10}\) & \(4.42\substack{+0.01 \\ -0.02}\) & \(0.153\substack{+0.007 \\ -0.007}\) & \(0.87\substack{+0.04 \\ -0.03}\) & \(3.0\substack{+1.0 \\ -0.6}\) \\
    \NGC{362}\(^{\dagger\ast}\) & \(53\substack{+5 \\ -5}\) & \(0.98\substack{+0.02 \\ -0.02}\) & \(0.24\substack{+0.01 \\ -0.01}\) & \(0.8\substack{+0.2 \\ -0.1}\) & \(1.67\substack{+0.06 \\ -0.05}\) & \(0.00015\substack{+0.00009 \\ -0.00005}\) & \(2.6\substack{+0.2 \\ -0.2}\) & \(1.06\substack{+0.04 \\ -0.05}\) & \(1.56\substack{+0.06 \\ -0.05}\) & \(9.13\substack{+0.06 \\ -0.07}\) & \(0.44\substack{+0.02 \\ -0.02}\) & \(0.96\substack{+0.03 \\ -0.06}\) & \(-1.3\substack{+0.2 \\ -0.1}\) \\
    \NGC{1261} & \(59\substack{+9 \\ -8}\) & \(0.53\substack{+0.01 \\ -0.01}\) & \(0.44\substack{+0.02 \\ -0.02}\) & \(0.3\substack{+0.1 \\ -0.1}\) & \(2.22\substack{+0.08 \\ -0.08}\) & \(0.000008\substack{+0.000014 \\ -0.000006}\) & \(3.9\substack{+0.2 \\ -0.2}\) & \(1.11\substack{+0.07 \\ -0.08}\) & \(1.69\substack{+0.08 \\ -0.07}\) & \(16.33\substack{+0.06 \\ -0.06}\) & \(0.46\substack{+0.02 \\ -0.02}\) & \(0.96\substack{+0.03 \\ -0.04}\) & \(-1.2\substack{+0.1 \\ -0.1}\) \\
    \NGC{1851} & \(61\substack{+5 \\ -6}\) & \(0.93\substack{+0.02 \\ -0.02}\) & \(0.32\substack{+0.01 \\ -0.01}\) & \(5\substack{+2 \\ -2}\) & \(1.96\substack{+0.02 \\ -0.02}\) & \(0.00002\substack{+0.00002 \\ -0.00001}\) & \(3.3\substack{+0.2 \\ -0.2}\) & \(0.76\substack{+0.05 \\ -0.05}\) & \(2.18\substack{+0.05 \\ -0.05}\) & \(12.39\substack{+0.07 \\ -0.08}\) & \(0.43\substack{+0.01 \\ -0.01}\) & \(0.96\substack{+0.03 \\ -0.05}\) & \(-2\substack{+2 \\ -2}\) \\
    \NGC{2808} & \(49\substack{+9 \\ -6}\) & \(2.02\substack{+0.02 \\ -0.03}\) & \(0.48\substack{+0.01 \\ -0.01}\) & \(1.8\substack{+0.2 \\ -0.3}\) & \(2.00\substack{+0.02 \\ -0.02}\) & \(0.000022\substack{+0.000011 \\ -0.000006}\) & \(4.82\substack{+0.05 \\ -0.05}\) & \(0.37\substack{+0.02 \\ -0.02}\) & \(2.84\substack{+0.03 \\ -0.03}\) & \(10.36\substack{+0.05 \\ -0.04}\) & \(0.41\substack{+0.01 \\ -0.01}\) & \(0.95\substack{+0.04 \\ -0.06}\) & \(-0.2\substack{+0.2 \\ -0.3}\) \\
    \NGC{4372} & \(8\substack{+5 \\ -2}\) & \(0.55\substack{+0.02 \\ -0.02}\) & \(0.39\substack{+0.04 \\ -0.03}\) & \(5\substack{+2 \\ -2}\) & \(1.6\substack{+0.3 \\ -0.2}\) & \(7\substack{+5 \\ -5}\) & \(1.43\substack{+0.12 \\ -0.10}\) & \(0.70\substack{+0.04 \\ -0.04}\) & \(1.24\substack{+0.10 \\ -0.10}\) & \(5.29\substack{+0.08 \\ -0.08}\) & \(0.20\substack{+0.08 \\ -0.06}\) & \(0.7\substack{+0.2 \\ -0.2}\) & \(-2\substack{+3 \\ -2}\) \\
    \NGC{4590} & \(49\substack{+12 \\ -10}\) & \(0.288\substack{+0.008 \\ -0.009}\) & \(0.50\substack{+0.02 \\ -0.02}\) & \(0.9\substack{+0.2 \\ -0.3}\) & \(1.7\substack{+0.2 \\ -0.2}\) & \(0.002\substack{+0.201 \\ -0.002}\) & \(2.2\substack{+0.2 \\ -0.2}\) & \(1.26\substack{+0.04 \\ -0.04}\) & \(1.61\substack{+0.08 \\ -0.06}\) & \(10.30\substack{+0.07 \\ -0.09}\) & \(0.44\substack{+0.03 \\ -0.03}\) & \(0.93\substack{+0.05 \\ -0.07}\) & \(-0.2\substack{+0.2 \\ -0.4}\) \\
    \NGC{4833} & \(31\substack{+4 \\ -4}\) & \(0.83\substack{+0.02 \\ -0.02}\) & \(0.45\substack{+0.03 \\ -0.03}\) & \(0.7\substack{+0.2 \\ -0.2}\) & \(0.8\substack{+0.1 \\ -0.1}\) & \(6\substack{+5 \\ -4}\) & \(1.9\substack{+0.2 \\ -0.1}\) & \(0.81\substack{+0.04 \\ -0.03}\) & \(1.42\substack{+0.06 \\ -0.07}\) & \(6.47\substack{+0.05 \\ -0.05}\) & \(0.32\substack{+0.01 \\ -0.02}\) & \(0.97\substack{+0.02 \\ -0.04}\) & \(-0.7\substack{+0.2 \\ -0.1}\) \\
    \NGC{5024} & \(9.9\substack{+0.5 \\ -0.4}\) & \(1.21\substack{+0.03 \\ -0.03}\) & \(1.02\substack{+0.08 \\ -0.06}\) & \(6\substack{+2 \\ -3}\) & \(2.0\substack{+0.1 \\ -0.1}\) & \(0.0007\substack{+0.0021 \\ -0.0006}\) & \(3.0\substack{+0.2 \\ -0.2}\) & \(0.94\substack{+0.08 \\ -0.07}\) & \(1.87\substack{+0.07 \\ -0.08}\) & \(18.9\substack{+0.1 \\ -0.1}\) & \(0.110\substack{+0.009 \\ -0.007}\) & \(0.98\substack{+0.01 \\ -0.03}\) & \(-0.3\substack{+1.9 \\ -2.3}\) \\
    \NGC{5139} & \(12.6\substack{+0.4 \\ -0.6}\) & \(6.81\substack{+0.04 \\ -0.03}\) & \(1.25\substack{+0.02 \\ -0.02}\) & \(0.89\substack{+0.04 \\ -0.06}\) & \(1.92\substack{+0.03 \\ -0.02}\) & \(0.0004\substack{+0.0001 \\ -0.0001}\) & \(4.97\substack{+0.02 \\ -0.03}\) & \(0.68\substack{+0.03 \\ -0.03}\) & \(1.53\substack{+0.01 \\ -0.02}\) & \(5.364\substack{+0.010 \\ -0.007}\) & \(0.44\substack{+0.02 \\ -0.02}\) & \(0.85\substack{+0.04 \\ -0.05}\) & \(0.20\substack{+0.03 \\ -0.03}\) \\
    \NGC{5272} & \(10.7\substack{+0.9 \\ -0.5}\) & \(1.24\substack{+0.02 \\ -0.02}\) & \(0.46\substack{+0.01 \\ -0.01}\) & \(6\substack{+2 \\ -2}\) & \(1.64\substack{+0.04 \\ -0.04}\) & \(0.000011\substack{+0.000038 \\ -0.000008}\) & \(2.3\substack{+0.1 \\ -0.1}\) & \(1.15\substack{+0.04 \\ -0.04}\) & \(1.41\substack{+0.05 \\ -0.05}\) & \(10.27\substack{+0.06 \\ -0.06}\) & \(0.115\substack{+0.013 \\ -0.009}\) & \(0.94\substack{+0.04 \\ -0.07}\) & \(0.3\substack{+1.9 \\ -2.2}\) \\
    \NGC{5904} & \(23\substack{+3 \\ -3}\) & \(1.17\substack{+0.03 \\ -0.03}\) & \(0.196\substack{+0.009 \\ -0.009}\) & \(5\substack{+2 \\ -2}\) & \(1.58\substack{+0.06 \\ -0.06}\) & \(0.005\substack{+0.011 \\ -0.003}\) & \(4.8\substack{+0.2 \\ -0.2}\) & \(0.66\substack{+0.05 \\ -0.05}\) & \(1.00\substack{+0.05 \\ -0.06}\) & \(7.33\substack{+0.05 \\ -0.05}\) & \(0.27\substack{+0.01 \\ -0.02}\) & \(0.98\substack{+0.01 \\ -0.03}\) & \(-1.0\substack{+2.1 \\ -2.3}\) \\
    \NGC{5986}\(^\dagger\) & \(13\substack{+4 \\ -3}\) & \(1.06\substack{+0.03 \\ -0.03}\) & \(0.71\substack{+0.07 \\ -0.06}\) & \(0.6\substack{+5.8 \\ -0.3}\) & \(2.0\substack{+0.2 \\ -0.1}\) & \(0.00010\substack{+0.00010 \\ -0.00007}\) & \(4.4\substack{+0.3 \\ -0.4}\) & \(1.20\substack{+0.07 \\ -0.06}\) & \(1.78\substack{+0.09 \\ -0.11}\) & \(10.45\substack{+0.08 \\ -0.09}\) & \(0.29\substack{+0.05 \\ -0.04}\) & \(0.96\substack{+0.03 \\ -0.06}\) & \(-0.7\substack{+0.8 \\ -1.4}\) \\
    \NGC{6093}\(^\dagger\) & \(22\substack{+11 \\ -5}\) & \(1.28\substack{+0.04 \\ -0.04}\) & \(0.056\substack{+0.005 \\ -0.004}\) & \(3\substack{+2 \\ -1}\) & \(1.37\substack{+0.05 \\ -0.04}\) & \(0.009\substack{+0.003 \\ -0.003}\) & \(4.5\substack{+0.4 \\ -0.4}\) & \(0.76\substack{+0.07 \\ -0.06}\) & \(2.01\substack{+0.07 \\ -0.07}\) & \(9.93\substack{+0.07 \\ -0.07}\) & \(0.493\substack{+0.005 \\ -0.012}\) & \(0.3\substack{+0.1 \\ -0.1}\) & \(-0.7\substack{+1.4 \\ -2.0}\) \\
    \NGC{6121}\(^\dagger\) & \(13.7\substack{+0.9 \\ -1.0}\) & \(0.76\substack{+0.01 \\ -0.01}\) & \(1.7\substack{+0.2 \\ -0.2}\) & \(5\substack{+2 \\ -2}\) & \(1.0\substack{+0.1 \\ -0.1}\) & \(0.00002\substack{+0.00006 \\ -0.00001}\) & \(1.7\substack{+0.1 \\ -0.1}\) & \(0.95\substack{+0.08 \\ -0.08}\) & \(1.70\substack{+0.09 \\ -0.10}\) & \(1.83\substack{+0.01 \\ -0.01}\) & \(0.15\substack{+0.01 \\ -0.01}\) & \(0.97\substack{+0.02 \\ -0.04}\) & \(-1\substack{+3 \\ -2}\) \\
    \NGC{6171} & \(23\substack{+4 \\ -3}\) & \(0.437\substack{+0.010 \\ -0.008}\) & \(0.88\substack{+0.07 \\ -0.07}\) & \(5\substack{+2 \\ -3}\) & \(0.8\substack{+0.1 \\ -0.1}\) & \(0.004\substack{+0.002 \\ -0.002}\) & \(2.3\substack{+0.2 \\ -0.1}\) & \(1.12\substack{+0.04 \\ -0.06}\) & \(1.20\substack{+0.07 \\ -0.05}\) & \(5.48\substack{+0.05 \\ -0.06}\) & \(0.28\substack{+0.02 \\ -0.02}\) & \(0.98\substack{+0.02 \\ -0.03}\) & \(-0.8\substack{+1.5 \\ -2.3}\) \\
    \NGC{6205} & \(24\substack{+7 \\ -8}\) & \(1.42\substack{+0.03 \\ -0.03}\) & \(0.164\substack{+0.011 \\ -0.010}\) & \(2.8\substack{+1.2 \\ -0.9}\) & \(2.55\substack{+0.04 \\ -0.03}\) & \(0.00002\substack{+0.00003 \\ -0.00001}\) & \(2.8\substack{+0.3 \\ -0.3}\) & \(0.56\substack{+0.08 \\ -0.09}\) & \(1.39\substack{+0.07 \\ -0.07}\) & \(7.38\substack{+0.03 \\ -0.04}\) & \(0.43\substack{+0.04 \\ -0.07}\) & \(0.7\substack{+0.1 \\ -0.2}\) & \(-0.7\substack{+1.1 \\ -1.3}\) \\
    \NGC{6218} & \(38\substack{+8 \\ -11}\) & \(0.46\substack{+0.01 \\ -0.01}\) & \(0.180\substack{+0.010 \\ -0.009}\) & \(4\substack{+3 \\ -3}\) & \(0.78\substack{+0.05 \\ -0.07}\) & \(0.00009\substack{+0.00024 \\ -0.00007}\) & \(3.0\substack{+0.3 \\ -0.3}\) & \(0.62\substack{+0.06 \\ -0.06}\) & \(1.36\substack{+0.07 \\ -0.06}\) & \(5.09\substack{+0.03 \\ -0.03}\) & \(0.495\substack{+0.004 \\ -0.009}\) & \(0.57\substack{+0.12 \\ -0.07}\) & \(-1\substack{+1 \\ -2}\) \\
    \NGC{6254} & \(32\substack{+6 \\ -6}\) & \(0.73\substack{+0.02 \\ -0.02}\) & \(0.25\substack{+0.01 \\ -0.01}\) & \(5\substack{+2 \\ -3}\) & \(0.99\substack{+0.07 \\ -0.08}\) & \(0.0003\substack{+0.0003 \\ -0.0002}\) & \(2.9\substack{+0.2 \\ -0.2}\) & \(0.69\substack{+0.04 \\ -0.04}\) & \(1.31\substack{+0.06 \\ -0.06}\) & \(5.06\substack{+0.04 \\ -0.04}\) & \(0.32\substack{+0.02 \\ -0.02}\) & \(0.95\substack{+0.03 \\ -0.07}\) & \(-1\substack{+2 \\ -2}\) \\
    \NGC{6266}\(^{\dagger\ast}\) & \(57\substack{+4 \\ -5}\) & \(2.43\substack{+0.07 \\ -0.05}\) & \(0.089\substack{+0.011 \\ -0.008}\) & \(1.6\substack{+3.1 \\ -0.5}\) & \(1.10\substack{+0.07 \\ -0.08}\) & \(0.24\substack{+0.03 \\ -0.02}\) & \(3.0\substack{+0.3 \\ -0.3}\) & \(0.05\substack{+0.16 \\ -0.18}\) & \(1.92\substack{+0.08 \\ -0.08}\) & \(6.36\substack{+0.04 \\ -0.04}\) & \(0.48\substack{+0.02 \\ -0.03}\) & \(0.97\substack{+0.02 \\ -0.03}\) & \(-1.2\substack{+0.8 \\ -0.5}\) \\
    \NGC{6341} & \(23\substack{+4 \\ -4}\) & \(0.92\substack{+0.03 \\ -0.02}\) & \(0.23\substack{+0.01 \\ -0.01}\) & \(1.8\substack{+0.9 \\ -0.6}\) & \(1.73\substack{+0.07 \\ -0.08}\) & \(0.00006\substack{+0.00008 \\ -0.00003}\) & \(4.0\substack{+0.2 \\ -0.2}\) & \(1.06\substack{+0.05 \\ -0.05}\) & \(1.51\substack{+0.07 \\ -0.07}\) & \(8.35\substack{+0.04 \\ -0.05}\) & \(0.36\substack{+0.03 \\ -0.04}\) & \(0.97\substack{+0.02 \\ -0.04}\) & \(-0.9\substack{+0.9 \\ -1.1}\) \\
    \NGC{6352} & \(26\substack{+2 \\ -2}\) & \(0.32\substack{+0.01 \\ -0.01}\) & \(0.51\substack{+0.05 \\ -0.04}\) & \(6\substack{+2 \\ -3}\) & \(0.04\substack{+0.06 \\ -0.03}\) & \(0.28\substack{+0.06 \\ -0.05}\) & \(2.9\substack{+0.3 \\ -0.2}\) & \(0.74\substack{+0.06 \\ -0.07}\) & \(0.82\substack{+0.07 \\ -0.06}\) & \(5.51\substack{+0.06 \\ -0.06}\) & \(0.229\substack{+0.008 \\ -0.009}\) & \(0.98\substack{+0.01 \\ -0.03}\) & \(-2\substack{+4 \\ -2}\) \\
    \NGC{6362} & \(10.8\substack{+0.9 \\ -5.2}\) & \(0.465\substack{+0.008 \\ -0.028}\) & \(0.37\substack{+0.30 \\ -0.02}\) & \(6.4\substack{+0.7 \\ -3.3}\) & \(1.93\substack{+0.08 \\ -1.56}\) & \(0.0004\substack{+13.9934 \\ -0.0004}\) & \(4.0\substack{+0.2 \\ -2.0}\) & \(0.647\substack{+0.156 \\ -0.007}\) & \(0.76\substack{+0.23 \\ -0.02}\) & \(7.56\substack{+0.08 \\ -0.09}\) & \(0.22\substack{+0.07 \\ -0.06}\) & \(0.81\substack{+0.16 \\ -0.01}\) & \(-0.7\substack{+0.9 \\ -1.8}\) \\
    \NGC{6366} & \(10\substack{+12 \\ -6}\) & \(0.262\substack{+0.010 \\ -0.009}\) & \(0.9\substack{+0.1 \\ -0.1}\) & \(5\substack{+2 \\ -3}\) & \(2.1\substack{+0.5 \\ -0.4}\) & \(0.002\substack{+0.002 \\ -0.001}\) & \(1.7\substack{+0.1 \\ -0.1}\) & \(0.60\substack{+0.09 \\ -0.10}\) & \(1.1\substack{+0.1 \\ -0.1}\) & \(3.47\substack{+0.04 \\ -0.04}\) & \(0.36\substack{+0.08 \\ -0.09}\) & \(0.89\substack{+0.09 \\ -0.18}\) & \(-0.9\substack{+2.6 \\ -2.7}\) \\
    \NGC{6397}\(^\ast\) & \(65\substack{+3 \\ -2}\) & \(0.417\substack{+0.007 \\ -0.008}\) & \(0.091\substack{+0.002 \\ -0.002}\) & \(5\substack{+2 \\ -3}\) & \(1.51\substack{+0.03 \\ -0.04}\) & \(1.0\substack{+0.5 \\ -0.4}\) & \(2.8\substack{+0.2 \\ -0.1}\) & \(1.14\substack{+0.04 \\ -0.04}\) & \(1.29\substack{+0.04 \\ -0.04}\) & \(2.39\substack{+0.01 \\ -0.01}\) & \(0.4990\substack{+0.0007 \\ -0.0020}\) & \(0.98\substack{+0.01 \\ -0.02}\) & \(-0.8\substack{+2.1 \\ -1.9}\) \\
    \NGC{6541}\(^\ast\) & \(60\substack{+4 \\ -5}\) & \(0.77\substack{+0.02 \\ -0.02}\) & \(0.186\substack{+0.008 \\ -0.008}\) & \(0.9\substack{+0.3 \\ -0.2}\) & \(1.15\substack{+0.08 \\ -0.06}\) & \(0.0004\substack{+0.0003 \\ -0.0002}\) & \(2.8\substack{+0.2 \\ -0.2}\) & \(0.84\substack{+0.05 \\ -0.05}\) & \(1.73\substack{+0.06 \\ -0.06}\) & \(7.32\substack{+0.05 \\ -0.06}\) & \(0.48\substack{+0.01 \\ -0.02}\) & \(0.95\substack{+0.04 \\ -0.07}\) & \(-1.5\substack{+0.2 \\ -0.2}\) \\
    \NGC{6681}\(^\dagger\) & \(50\substack{+4 \\ -4}\) & \(0.60\substack{+0.02 \\ -0.01}\) & \(0.20\substack{+0.02 \\ -0.02}\) & \(1.1\substack{+0.1 \\ -0.1}\) & \(0.94\substack{+0.08 \\ -0.08}\) & \(0.03\substack{+0.02 \\ -0.01}\) & \(2.5\substack{+0.2 \\ -0.2}\) & \(1.06\substack{+0.03 \\ -0.04}\) & \(1.13\substack{+0.04 \\ -0.03}\) & \(9.17\substack{+0.09 \\ -0.08}\) & \(0.44\substack{+0.02 \\ -0.02}\) & \(0.989\substack{+0.008 \\ -0.018}\) & \(-1.5\substack{+0.2 \\ -0.2}\) \\
    \NGC{6723} & \(20\substack{+4 \\ -5}\) & \(0.73\substack{+0.01 \\ -0.01}\) & \(0.23\substack{+0.01 \\ -0.01}\) & \(0.08\substack{+0.10 \\ -0.08}\) & \(2.20\substack{+0.10 \\ -0.11}\) & \(0.005\substack{+0.004 \\ -0.003}\) & \(2.3\substack{+0.2 \\ -0.2}\) & \(0.75\substack{+0.08 \\ -0.08}\) & \(1.02\substack{+0.05 \\ -0.06}\) & \(8.00\substack{+0.08 \\ -0.07}\) & \(0.43\substack{+0.04 \\ -0.04}\) & \(0.86\substack{+0.09 \\ -0.11}\) & \(-1.0\substack{+0.1 \\ -0.1}\) \\
    \NGC{6752}\(^\ast\) & \(69\substack{+5 \\ -4}\) & \(0.96\substack{+0.02 \\ -0.02}\) & \(0.062\substack{+0.004 \\ -0.004}\) & \(7.7\substack{+0.2 \\ -0.4}\) & \(1.63\substack{+0.04 \\ -0.04}\) & \(14.4\substack{+0.5 \\ -0.8}\) & \(4.2\substack{+0.2 \\ -0.2}\) & \(0.93\substack{+0.04 \\ -0.06}\) & \(1.04\substack{+0.05 \\ -0.04}\) & \(4.15\substack{+0.01 \\ -0.01}\) & \(0.495\substack{+0.004 \\ -0.009}\) & \(0.90\substack{+0.06 \\ -0.08}\) & \(2.0\substack{+0.6 \\ -0.7}\) \\
    \NGC{6779} & \(39\substack{+8 \\ -6}\) & \(0.541\substack{+0.009 \\ -0.010}\) & \(0.63\substack{+0.04 \\ -0.04}\) & \(6\substack{+1 \\ -3}\) & \(0.96\substack{+0.08 \\ -0.09}\) & \(0.0006\substack{+0.0004 \\ -0.0003}\) & \(2.7\substack{+0.3 \\ -0.3}\) & \(0.93\substack{+0.04 \\ -0.04}\) & \(1.68\substack{+0.06 \\ -0.06}\) & \(10.49\substack{+0.07 \\ -0.06}\) & \(0.28\substack{+0.01 \\ -0.02}\) & \(0.90\substack{+0.08 \\ -0.11}\) & \(-0.8\substack{+2.1 \\ -2.2}\) \\
    \NGC{6809} & \(3\substack{+2 \\ -1}\) & \(0.74\substack{+0.03 \\ -0.02}\) & \(0.30\substack{+0.02 \\ -0.02}\) & \(5\substack{+2 \\ -2}\) & \(2.3\substack{+0.2 \\ -0.2}\) & \(0.0001\substack{+0.0002 \\ -0.0001}\) & \(4.5\substack{+0.3 \\ -0.3}\) & \(0.81\substack{+0.06 \\ -0.06}\) & \(1.06\substack{+0.09 \\ -0.08}\) & \(5.31\substack{+0.04 \\ -0.04}\) & \(0.30\substack{+0.06 \\ -0.05}\) & \(0.91\substack{+0.07 \\ -0.13}\) & \(0.5\substack{+2.5 \\ -2.8}\) \\
    \NGC{7078}\(^\ast\) & \(25\substack{+1 \\ -1}\) & \(1.64\substack{+0.03 \\ -0.02}\) & \(0.24\substack{+0.01 \\ -0.01}\) & \(1.4\substack{+0.1 \\ -0.1}\) & \(1.32\substack{+0.05 \\ -0.06}\) & \(0.00006\substack{+0.00005 \\ -0.00004}\) & \(3.8\substack{+0.3 \\ -0.3}\) & \(0.82\substack{+0.06 \\ -0.06}\) & \(1.59\substack{+0.05 \\ -0.05}\) & \(10.67\substack{+0.06 \\ -0.08}\) & \(0.23\substack{+0.01 \\ -0.02}\) & \(0.88\substack{+0.07 \\ -0.09}\) & \(-0.4\substack{+0.1 \\ -0.1}\) \\
    \NGC{7089} & \(38\substack{+8 \\ -7}\) & \(1.75\substack{+0.04 \\ -0.05}\) & \(0.28\substack{+0.03 \\ -0.02}\) & \(5\substack{+2 \\ -2}\) & \(2.12\substack{+0.08 \\ -0.08}\) & \(0.00004\substack{+0.00004 \\ -0.00002}\) & \(4.2\substack{+0.4 \\ -0.4}\) & \(0.3\substack{+0.2 \\ -0.2}\) & \(1.99\substack{+0.10 \\ -0.09}\) & \(11.63\substack{+0.07 \\ -0.08}\) & \(0.36\substack{+0.02 \\ -0.03}\) & \(0.94\substack{+0.05 \\ -0.09}\) & \(-2\substack{+2 \\ -2}\) \\
    \NGC{7099}\(^\ast\) & \(47\substack{+2 \\ -1}\) & \(0.526\substack{+0.007 \\ -0.011}\) & \(0.269\substack{+0.016 \\ -0.005}\) & \(1.15\substack{+0.09 \\ -0.08}\) & \(0.77\substack{+0.06 \\ -0.06}\) & \(0.2\substack{+0.4 \\ -0.2}\) & \(3.9\substack{+0.3 \\ -0.3}\) & \(1.00\substack{+0.04 \\ -0.03}\) & \(1.42\substack{+0.07 \\ -0.04}\) & \(8.50\substack{+0.10 \\ -0.08}\) & \(0.349\substack{+0.008 \\ -0.007}\) & \(0.993\substack{+0.005 \\ -0.007}\) & \(-1.30\substack{+0.14 \\ -0.07}\) \\
    \bottomrule
    \end{tabular}
    \end{rotatetable}
    \caption{
        Median and \(1\sigma\) uncertainties of the best-fitting
        model parameters, for all clusters in our sample.
        Note that the uncertainties presented here represent only the
        statistical uncertainties, and likely underestimate the true errors
        All clusters classified as core-collapsed in \citet{Trager1995}
        are denoted by an asterisk. All clusters with
        \(\RGeff<\SI{1.5}{\kilo\pc}\) are denoted by a dagger.
    }
    \label{table:best_fitting_params}
    \end{table}

\begin{table}
    \renewcommand*{\arraystretch}{1.3}
    \centering
    \scriptsize
    \begin{tabular}{lccccccc}
    \toprule
    \multirow{3}{*}{Cluster}    & \multicolumn{3}{c}{Initial Conditions} & \multicolumn{4}{c}{Present Day Conditions} \\
            \cmidrule(l){2-4} \cmidrule(l){5-8}
     & \(\log_{10}\left( \rho_{\mathrm{h},0} \right)\) & \(\mathrm{f}_{\mathrm{BH},0}\) & \(v_{\mathrm{esc},0}\) & \(\log_{10}\left( \rho_{\mathrm{h}} \right)\) & \(f_{\mathrm{BH}}\) & \(M\) & \(\rh\)\\
     & \(\left[M_\odot\ \mathrm{pc^{-3}}\right]\) & \(\left[\%\right]\) & \(\left[\mathrm{km}/\mathrm{s}\right]\) & \(\left[M_\odot\ \mathrm{pc^{-3}}\right]\) & \(\left[\%\right]\) & \(\left[10^6\,\Msun\right]\) & \(\left[\mathrm{pc}\right]\)\\
    \midrule
    \NGC{104} & \(6.01\substack{+0.14 \\ -0.07}\) & \(5.44\substack{+0.15 \\ -0.08}\) & \(196\substack{+13 \\ -7}\) & \(2.78\substack{+0.02 \\ -0.01}\) & \(1.23\substack{+0.03 \\ -0.02}\) & \(0.810\substack{+0.007 \\ -0.006}\) & \(5.45\substack{+0.04 \\ -0.05}\) \\
    \NGC{362}\(^{\dagger\ast}\) & \(6.94\substack{+0.07 \\ -0.07}\) & \(7.07\substack{+0.07 \\ -0.08}\) & \(225\substack{+7 \\ -8}\) & \(2.96\substack{+0.02 \\ -0.02}\) & \(0.30\substack{+0.01 \\ -0.02}\) & \(0.297\substack{+0.005 \\ -0.004}\) & \(3.40\substack{+0.04 \\ -0.05}\) \\
    \NGC{1261} & \(5.87\substack{+0.07 \\ -0.07}\) & \(6.41\substack{+0.08 \\ -0.09}\) & \(122\substack{+4 \\ -4}\) & \(2.25\substack{+0.01 \\ -0.01}\) & \(0.30\substack{+0.02 \\ -0.02}\) & \(0.188\substack{+0.003 \\ -0.003}\) & \(5.03\substack{+0.05 \\ -0.05}\) \\
    \NGC{1851} & \(6.54\substack{+0.06 \\ -0.06}\) & \(6.20\substack{+0.07 \\ -0.07}\) & \(190\substack{+5 \\ -5}\) & \(2.97\substack{+0.01 \\ -0.01}\) & \(0.21\substack{+0.01 \\ -0.01}\) & \(0.329\substack{+0.004 \\ -0.004}\) & \(3.49\substack{+0.03 \\ -0.03}\) \\
    \NGC{2808} & \(6.34\substack{+0.04 \\ -0.04}\) & \(5.58\substack{+0.05 \\ -0.04}\) & \(227\substack{+4 \\ -4}\) & \(3.331\substack{+0.010 \\ -0.007}\) & \(0.461\substack{+0.008 \\ -0.009}\) & \(0.931\substack{+0.008 \\ -0.010}\) & \(3.73\substack{+0.01 \\ -0.02}\) \\
    \NGC{3201} & \(6.7\substack{+0.2 \\ -0.1}\) & \(7.9\substack{+0.2 \\ -0.1}\) & \(158\substack{+11 \\ -9}\) & \(1.79\substack{+0.03 \\ -0.03}\) & \(0.37\substack{+0.03 \\ -0.04}\) & \(0.142\substack{+0.003 \\ -0.003}\) & \(6.5\substack{+0.2 \\ -0.2}\) \\
    \NGC{4372} & \(6.0\substack{+0.1 \\ -0.1}\) & \(7.8\substack{+0.2 \\ -0.2}\) & \(131\substack{+8 \\ -8}\) & \(1.67\substack{+0.02 \\ -0.02}\) & \(1.2\substack{+0.1 \\ -0.1}\) & \(0.163\substack{+0.005 \\ -0.005}\) & \(7.4\substack{+0.1 \\ -0.1}\) \\
    \NGC{4590} & \(5.44\substack{+0.07 \\ -0.06}\) & \(6.56\substack{+0.08 \\ -0.08}\) & \(84\substack{+3 \\ -3}\) & \(1.61\substack{+0.02 \\ -0.02}\) & \(0.20\substack{+0.02 \\ -0.02}\) & \(0.116\substack{+0.003 \\ -0.003}\) & \(7.0\substack{+0.1 \\ -0.1}\) \\
    \NGC{4833} & \(6.05\substack{+0.10 \\ -0.08}\) & \(7.57\substack{+0.13 \\ -0.08}\) & \(151\substack{+8 \\ -5}\) & \(2.37\substack{+0.01 \\ -0.01}\) & \(1.09\substack{+0.05 \\ -0.06}\) & \(0.188\substack{+0.004 \\ -0.003}\) & \(4.58\substack{+0.05 \\ -0.05}\) \\
    \NGC{5024} & \(5.14\substack{+0.09 \\ -0.10}\) & \(6.7\substack{+0.1 \\ -0.1}\) & \(121\substack{+5 \\ -5}\) & \(1.82\substack{+0.02 \\ -0.01}\) & \(2.8\substack{+0.1 \\ -0.1}\) & \(0.55\substack{+0.01 \\ -0.01}\) & \(9.9\substack{+0.2 \\ -0.1}\) \\
    \NGC{5139} & \(5.62\substack{+0.02 \\ -0.02}\) & \(7.85\substack{+0.03 \\ -0.03}\) & \(259\substack{+2 \\ -2}\) & \(2.608\substack{+0.003 \\ -0.002}\) & \(8.31\substack{+0.06 \\ -0.06}\) & \(3.02\substack{+0.01 \\ -0.01}\) & \(9.62\substack{+0.03 \\ -0.02}\) \\
    \NGC{5272} & \(6.19\substack{+0.05 \\ -0.04}\) & \(7.30\substack{+0.07 \\ -0.06}\) & \(183\substack{+4 \\ -4}\) & \(2.21\substack{+0.01 \\ -0.01}\) & \(1.54\substack{+0.08 \\ -0.07}\) & \(0.507\substack{+0.007 \\ -0.008}\) & \(7.18\substack{+0.09 \\ -0.08}\) \\
    \NGC{5904} & \(7.27\substack{+0.07 \\ -0.07}\) & \(8.26\substack{+0.07 \\ -0.08}\) & \(271\substack{+9 \\ -9}\) & \(2.28\substack{+0.02 \\ -0.02}\) & \(1.22\substack{+0.07 \\ -0.07}\) & \(0.369\substack{+0.007 \\ -0.007}\) & \(6.13\substack{+0.09 \\ -0.09}\) \\
    \NGC{5986}\(^\dagger\) & \(5.5\substack{+0.1 \\ -0.1}\) & \(6.6\substack{+0.2 \\ -0.1}\) & \(135\substack{+9 \\ -7}\) & \(2.72\substack{+0.02 \\ -0.02}\) & \(0.86\substack{+0.10 \\ -0.10}\) & \(0.284\substack{+0.007 \\ -0.006}\) & \(4.02\substack{+0.06 \\ -0.06}\) \\
    \NGC{6093}\(^\dagger\) & \(8.9\substack{+0.1 \\ -0.1}\) & \(7.60\substack{+0.05 \\ -0.05}\) & \(530\substack{+30 \\ -30}\) & \(3.385\substack{+0.010 \\ -0.009}\) & \(0.029\substack{+0.003 \\ -0.002}\) & \(0.292\substack{+0.005 \\ -0.004}\) & \(2.43\substack{+0.02 \\ -0.01}\) \\
    \NGC{6121}\(^\dagger\) & \(4.3\substack{+0.1 \\ -0.1}\) & \(5.97\substack{+0.09 \\ -0.11}\) & \(74\substack{+3 \\ -4}\) & \(2.33\substack{+0.01 \\ -0.01}\) & \(3.5\substack{+0.2 \\ -0.2}\) & \(0.084\substack{+0.002 \\ -0.001}\) & \(3.62\substack{+0.05 \\ -0.04}\) \\
    \NGC{6171} & \(4.9\substack{+0.1 \\ -0.1}\) & \(6.11\substack{+0.08 \\ -0.08}\) & \(78\substack{+4 \\ -3}\) & \(2.21\substack{+0.02 \\ -0.02}\) & \(0.86\substack{+0.08 \\ -0.07}\) & \(0.057\substack{+0.001 \\ -0.001}\) & \(3.47\substack{+0.05 \\ -0.04}\) \\
    \NGC{6205} & \(7.59\substack{+0.09 \\ -0.10}\) & \(8.27\substack{+0.11 \\ -0.10}\) & \(330\substack{+10 \\ -10}\) & \(2.741\substack{+0.009 \\ -0.009}\) & \(0.72\substack{+0.04 \\ -0.03}\) & \(0.436\substack{+0.005 \\ -0.006}\) & \(4.55\substack{+0.04 \\ -0.03}\) \\
    \NGC{6218} & \(6.98\substack{+0.07 \\ -0.09}\) & \(7.61\substack{+0.08 \\ -0.10}\) & \(178\substack{+7 \\ -8}\) & \(2.297\substack{+0.010 \\ -0.010}\) & \(0.162\substack{+0.013 \\ -0.009}\) & \(0.098\substack{+0.003 \\ -0.002}\) & \(3.88\substack{+0.05 \\ -0.03}\) \\
    \NGC{6254} & \(6.72\substack{+0.08 \\ -0.07}\) & \(7.80\substack{+0.10 \\ -0.08}\) & \(187\substack{+7 \\ -6}\) & \(2.25\substack{+0.02 \\ -0.02}\) & \(0.63\substack{+0.06 \\ -0.05}\) & \(0.205\substack{+0.004 \\ -0.004}\) & \(5.16\substack{+0.08 \\ -0.10}\) \\
    \NGC{6266}\(^{\dagger\ast}\) & \(8.6\substack{+0.1 \\ -0.2}\) & \(7.51\substack{+0.04 \\ -0.04}\) & \(580\substack{+30 \\ -40}\) & \(3.50\substack{+0.01 \\ -0.02}\) & \(0.25\substack{+0.04 \\ -0.02}\) & \(0.69\substack{+0.02 \\ -0.01}\) & \(2.95\substack{+0.06 \\ -0.03}\) \\
    \NGC{6341} & \(6.96\substack{+0.10 \\ -0.10}\) & \(7.7\substack{+0.1 \\ -0.1}\) & \(220\substack{+10 \\ -10}\) & \(2.67\substack{+0.02 \\ -0.02}\) & \(0.31\substack{+0.03 \\ -0.02}\) & \(0.294\substack{+0.006 \\ -0.006}\) & \(4.22\substack{+0.08 \\ -0.07}\) \\
    \NGC{6352} & \(5.5\substack{+0.1 \\ -0.1}\) & \(5.97\substack{+0.09 \\ -0.09}\) & \(89\substack{+6 \\ -5}\) & \(1.70\substack{+0.02 \\ -0.02}\) & \(1.10\substack{+0.08 \\ -0.08}\) & \(0.061\substack{+0.002 \\ -0.002}\) & \(5.26\substack{+0.08 \\ -0.08}\) \\
    \NGC{6362} & \(6.06\substack{+0.06 \\ -0.81}\) & \(6.96\substack{+0.04 \\ -0.46}\) & \(125\substack{+3 \\ -35}\) & \(1.71\substack{+0.01 \\ -0.18}\) & \(1.29\substack{+0.89 \\ -0.02}\) & \(0.097\substack{+0.003 \\ -0.002}\) & \(6.06\substack{+1.00 \\ -0.04}\) \\
    \NGC{6366} & \(4.7\substack{+0.2 \\ -0.2}\) & \(5.6\substack{+0.1 \\ -0.1}\) & \(61\substack{+5 \\ -5}\) & \(1.72\substack{+0.02 \\ -0.02}\) & \(1.3\substack{+0.1 \\ -0.1}\) & \(0.033\substack{+0.001 \\ -0.001}\) & \(4.23\substack{+0.06 \\ -0.07}\) \\
    \NGC{6397}\(^\ast\) & \(6.1\substack{+0.1 \\ -0.1}\) & \(7.73\substack{+0.12 \\ -0.10}\) & \(127\substack{+8 \\ -6}\) & \(1.84\substack{+0.03 \\ -0.03}\) & \(0.44\substack{+0.04 \\ -0.03}\) & \(0.115\substack{+0.002 \\ -0.002}\) & \(5.8\substack{+0.1 \\ -0.1}\) \\
    \NGC{6541}\(^\ast\) & \(7.16\substack{+0.07 \\ -0.06}\) & \(7.55\substack{+0.08 \\ -0.09}\) & \(226\substack{+8 \\ -6}\) & \(2.77\substack{+0.01 \\ -0.01}\) & \(0.160\substack{+0.011 \\ -0.009}\) & \(0.226\substack{+0.003 \\ -0.003}\) & \(3.59\substack{+0.04 \\ -0.04}\) \\
    \NGC{6681}\(^\dagger\) & \(6.9\substack{+0.1 \\ -0.1}\) & \(7.87\substack{+0.09 \\ -0.08}\) & \(189\substack{+12 \\ -10}\) & \(2.86\substack{+0.02 \\ -0.02}\) & \(0.15\substack{+0.02 \\ -0.01}\) & \(0.089\substack{+0.002 \\ -0.002}\) & \(2.45\substack{+0.04 \\ -0.03}\) \\
    \NGC{6723} & \(6.86\substack{+0.06 \\ -0.07}\) & \(7.36\substack{+0.05 \\ -0.06}\) & \(198\substack{+6 \\ -6}\) & \(2.29\substack{+0.02 \\ -0.01}\) & \(0.65\substack{+0.03 \\ -0.04}\) & \(0.169\substack{+0.003 \\ -0.003}\) & \(4.70\substack{+0.05 \\ -0.09}\) \\
    \NGC{6752}\(^\ast\) & \(8.69\substack{+0.09 \\ -0.07}\) & \(8.60\substack{+0.03 \\ -0.03}\) & \(440\substack{+20 \\ -10}\) & \(2.70\substack{+0.01 \\ -0.01}\) & \(0.124\substack{+0.012 \\ -0.010}\) & \(0.226\substack{+0.003 \\ -0.003}\) & \(3.77\substack{+0.04 \\ -0.04}\) \\
    \NGC{6779} & \(5.41\substack{+0.10 \\ -0.09}\) & \(6.93\substack{+0.09 \\ -0.09}\) & \(103\substack{+4 \\ -4}\) & \(2.03\substack{+0.02 \\ -0.02}\) & \(0.72\substack{+0.05 \\ -0.05}\) & \(0.151\substack{+0.002 \\ -0.002}\) & \(5.54\substack{+0.07 \\ -0.07}\) \\
    \NGC{6809} & \(6.5\substack{+0.1 \\ -0.1}\) & \(8.2\substack{+0.1 \\ -0.1}\) & \(173\substack{+9 \\ -9}\) & \(2.05\substack{+0.01 \\ -0.01}\) & \(1.12\substack{+0.07 \\ -0.06}\) & \(0.184\substack{+0.005 \\ -0.005}\) & \(5.83\substack{+0.06 \\ -0.05}\) \\
    \NGC{7078}\(^\ast\) & \(7.15\substack{+0.08 \\ -0.08}\) & \(7.99\substack{+0.11 \\ -0.09}\) & \(290\substack{+11 \\ -9}\) & \(2.710\substack{+0.010 \\ -0.012}\) & \(0.84\substack{+0.03 \\ -0.03}\) & \(0.620\substack{+0.007 \\ -0.008}\) & \(5.25\substack{+0.04 \\ -0.05}\) \\
    \NGC{7089} & \(7.0\substack{+0.1 \\ -0.1}\) & \(7.6\substack{+0.2 \\ -0.2}\) & \(280\substack{+20 \\ -20}\) & \(2.90\substack{+0.03 \\ -0.02}\) & \(0.90\substack{+0.07 \\ -0.09}\) & \(0.61\substack{+0.02 \\ -0.01}\) & \(4.5\substack{+0.1 \\ -0.1}\) \\
    \NGC{7099}\(^\ast\) & \(6.51\substack{+0.03 \\ -0.09}\) & \(7.51\substack{+0.05 \\ -0.13}\) & \(155\substack{+3 \\ -6}\) & \(2.336\substack{+0.007 \\ -0.008}\) & \(0.205\substack{+0.006 \\ -0.013}\) & \(0.140\substack{+0.002 \\ -0.002}\) & \(4.25\substack{+0.04 \\ -0.03}\) \\
    \bottomrule
    \end{tabular}
    \caption{
        Median and \(1\sigma\) uncertainties of certain derived quantities
        (i.e. parameters not directly varied during fitting) of
        the best-fitting models, for all clusters in our sample.
        These include the initial half-mass density, BH mass fraction and
        central escape velocity, and the
        present-day half-mass density, BH mass fraction, total cluster mass
        and half-mass radius. The initial mass and radius values
        are listed in \Cref{table:best_fitting_params}.
        Note that the uncertainties presented here represent only the
        statistical uncertainties, and likely underestimate the true errors
        All clusters classified as core-collapsed in \citet{Trager1995}
        are denoted by an asterisk. All clusters with
        \(\RGeff<\SI{1.5}{\kilo\pc}\) are denoted by a dagger.
    }
    \label{table:extra_params}
\end{table}

\section{CMC Mass Function Comparison}
\label{sec:cmc_massfunc_comp}

    \begin{figure*}
        \centering
        \includegraphics[width=\linewidth]{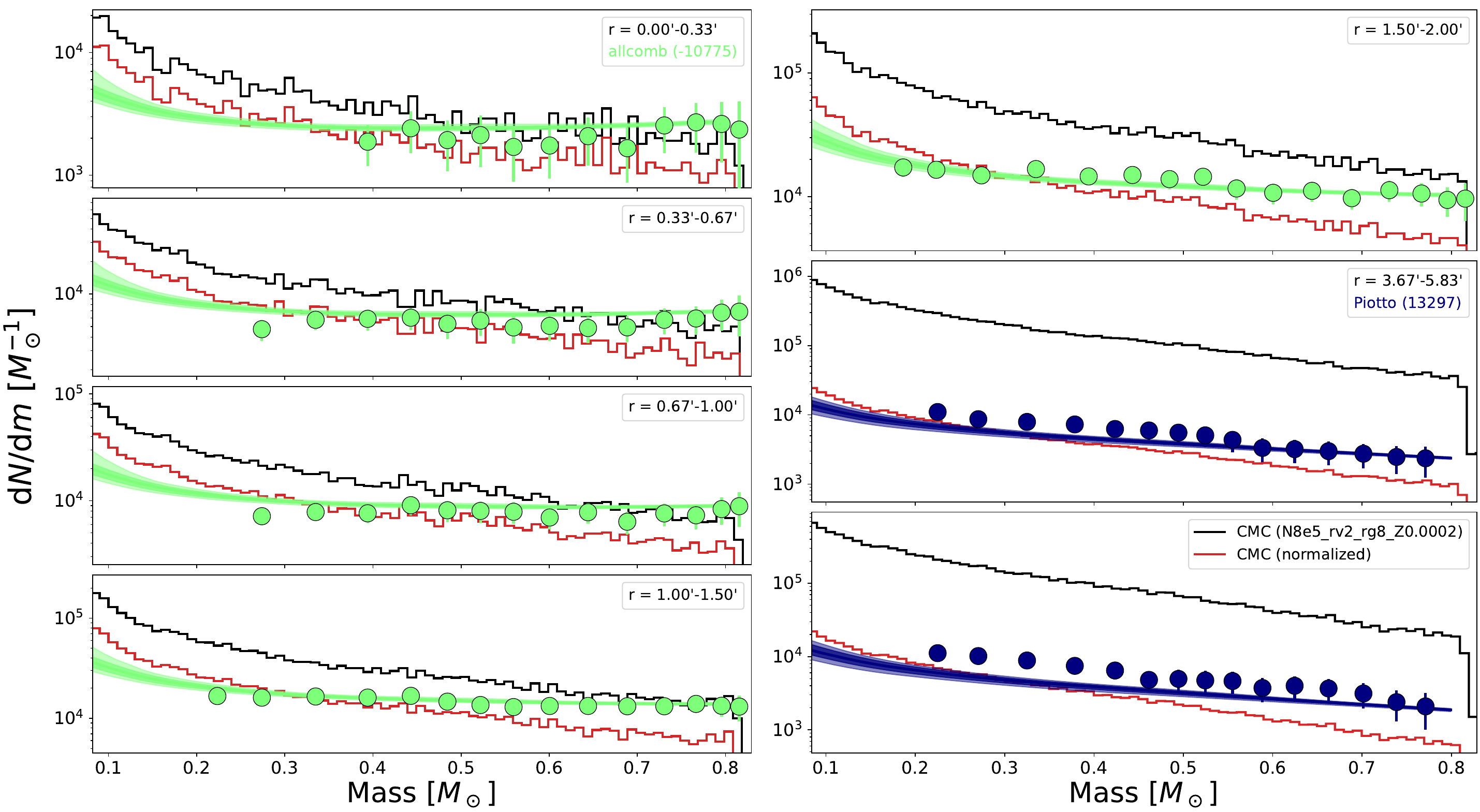}
        \caption{
            Comparison of the present-day local stellar mass functions
            between our fit to \NGC{3201} and the CMC model best matching
            the velocity dispersion and surface brightness profile of the cluster
            (\(N_0=8\times10^5\), \(r_{v,0}=\SI{2}{\pc}\),
            \(R_g=\SI{8}{\kilo\pc}\) and \(Z=0.0002\))
            \citep{Kremer2019,Rui2021b}.
            Each panel shows the number of stars per unit mass
            as a function of stellar mass, for different projected distance
            ranges from the cluster centre.
            The dark and light shaded regions represent the \(1\sigma\)
            and \(2\sigma\) credible intervals of the model fits, respectively.
            The measurements used to constrain the models are shown alongside
            their \(1\sigma\) uncertainties by the circles and errorbars.
            The original CMC snapshot mass functions are shown in black. In
            red, the same CMC mass functions are scaled to align with our fits,
            in order to highlight the difference in slope between the two.
        }
        \label{fig:cmc_massfunc_comp}
    \end{figure*}

    To demonstrate the origin of the IMF we infer
    (\Cref{sub:initial_mass_function}), in \Cref{fig:cmc_massfunc_comp} we
    compare our fits to an example cluster (\NGC{3201}) with the best-matching
    CMC model of this cluster.
    For this comparison, we use the CMC snapshot that has been used in previous
    studies to represent \NGC{3201} \citep[e.g.][]{Kremer2019,Vitral2022}, based
    on matching the CMC snapshots to the observed surface brightness and
    velocity dispersion profiles \citep{Rui2021b}
    \footnote{\citet{Rui2021b} also note that the same CMC model but with
    \(N_0=\SI{4e5}{\Msun}\) is also a good match under their methodology, but
    the differences shown here remain similar, and thus we opt to show the
    snapshot which is used in other studies.}.
    We also overplot, in red, the same CMC mass function scaled to match
    the mass functions of our best-fitting model within each panel, in order to
    more easily compare the slope of the mass function, irrespective of the
    total number of stars, which could likely be improved by choosing a
    lower-mass CMC model.

    It is clearly seen in \Cref{fig:cmc_massfunc_comp} that the CMC snapshot,
    which has been previously shown to match the density and kinematic data of
    \NGC{3201} well, is unable to reproduce the mass function data.
    The slopes of these local mass functions (and of the the global
    mass function) in CMC are too steep, and lead to too many low-mass stars.
    This is the direct consequence of the \citet{Kroupa2001} IMF
    assumed in the CMC grid.
    The present-day mass function datasets place useful constraints on the
    IMF, and are largely the reason we infer a bottom-light IMF across our
    sample.
    It would be valuable, in the future, to compute a similar grid of CMC
    simulations adopting a non-canonical IMF, and to re-examine which models
    best match all of the given data, and what impacts this would have on
    studies which use these models to represent MW clusters.


\end{document}